\documentclass[12pt]{article}
\usepackage{epsfig}
\begin{document}

\newcommand{\beq}{\begin{equation}}
\newcommand{\eeq}{\end{equation}}
\newcommand{\nn}{\nonumber}

\def\ii{\'{\i}}
\def\r{\rightarrow}
\def\err{\end{array}}
\def\bea{\begin{eqnarray}}
\def\eea{\end{eqnarray}}
\def\bp{{\bf p}}
\def\bk{{\bf k}}
\def\bq{{\bf q}}
\def\ttau{\tilde{\tau}}
\def\tchi{\tilde{\chi}}
\def\trho{\tilde{\rho}}
\def\teps{\tilde{\epsilon}}
\def\tnu{\tilde{\nu}}
\def\tgamma{\tilde{\gamma}}

\title{Solutions of the Polchinski ERG equation in the $O(N)$ scalar model}

\author{Yu.A. Kubyshin\footnote{On leave of absence from the Institute for Nuclear Physics, 
Moscow State University, 119899 Moscow, Russia.} \\
{\small \em Departament de Matem\`atica Aplicada IV, 
Universitat Polit\`ecnica de Catalunya} \\
{\small \em C-3, Campus Nord, C. Jordi Girona, 1-3, 
08034 Barcelona, Spain}\\
{\small \em yuri@mat.upc.es \vspace{0.2cm} }\\ 
\and
R. Neves and R. Potting \\
{\small \em \'Area Departamental de F{\ii}sica, FCT, 
Universidade do Algarve}\\
{\small \em Campus de Gambelas, 8000-117 Faro, Portugal}\\
{\small \em rneves@ualg.pt, rpotting@ualg.pt} 
} 

\date{}

\maketitle

\begin{abstract} 
\noindent Solutions of the Polchinski exact renormalization group equation in the 
scalar $O(N)$ theory are studied. 
Families of regular solutions are found and their relation with fixed 
points of the theory is established. Special attention is devoted 
to the limit $N = \infty$, where many properties can be 
analyzed analytically.
\vspace{0.25cm}

\end{abstract}

\section{Introduction}
\label{sect:Intr}

The Exact Renormalization Group (ERG) has proven to be a 
powerful non-perturba\-tive method for 
studies of phenomena in quantum field theories at various scales. It 
stems from the Wilson renormalization group \cite{W-WK} (see, for 
example, Ref.~\cite{Ma} for a review) and is, 
basically, its continuous version adapted to quantum 
field theory \cite{WH}-\cite{HH} 
(see Refs.\ \cite{Mo-rev}-\cite{IvLis} for reviews). 
The ERG method allows for general non-perturbative studies 
of low-energy effective actions, in particular their flows and 
fixed points, and is also recognized as a rather powerful calculational 
framework. 

The central object within the ERG approach is the (Wilsonian) effective 
action $S_{\Lambda}$ or the Legendre effective action, 
which gives an adequate 
description of the theory at the scale $\Lambda$ in terms of 
renormalized quantities. The action satisfies an equation in functional 
derivatives which determines completely the renormalization group 
evolution of $S_{\Lambda}$ with the scale. This equation is called the 
ERG equation.

Various formulations of the ERG have been proposed, the most widely 
used are the Wegner-Houghton sharp cutoff equation \cite{WH}, 
Polchinski ERG equation for the Wilson effective action 
\cite{Po} (see Ref.~\cite{BaTh} for detailed derivation and 
discussion), and equations for the Legendre effective action 
\cite{NiCh}-\cite{Mo}. All these equations are essentially
equivalent, the relation between them was clarified in 
Refs.\ \cite{Mo} and \cite{LaMo00}. It turns out that in practice they 
are too complicated to be solved exactly. Two well-developed 
non-perturbative approximate schemes for solving the ERG equations are: 
(1) truncation in the operator content to a finite set of operators 
\cite{polynom}, and (2) expansion in powers of space-time 
derivatives \cite{Myerson}-\cite{Mo94}. Whereas the first 
scheme is of limited reliability \cite{Mo94a,HKLM}, the second one, 
called the derivative expansion, was shown to be quite reliable and 
accurate enough in finding the fixed points and calculating the 
critical exponents \cite{HH,Go86,Mo94},\cite{NiChSt}-\cite{TeWe94}. 
In particular, the fixed point (FP) corresponding to the Ising model 
in three dimensions \cite{HH,Mo94,BHLM},\cite{Feld}-\cite{TeWe94}, 
and the conformal FPs in two dimensions \cite{Mo95,KNR} were 
calculated using the derivative expansion. The lowest order of 
the derivative expansion, ${\cal O}(\partial^{0})$-order, 
also called the local potential approximation (LPA), 
already gives rather good accuracy in a large variety of problems 
(see Refs.~\cite{HH} and \cite{NiChSt}). The next-to-leading order  
(${\cal O}(\partial^{2})$-order) usually improves the results of the 
leading order and has also been thoroughly studied in the literature 
(see, for example, Refs.~\cite{Go86,BHLM,Mo96-RG} and \cite{Co98}). 

To outline some general features of the ERG approach let us consider 
the case of a scalar field 
in $d$ dimensions. A generalization  
to the fermionic case is straightforward \cite{CKM}; the gauge invariant 
formalism is, however, more involved \cite{Mo99}.  
For a scalar field $\phi$ the effective interactions within the LPA are 
restricted to a potential $V_{\Lambda}(\phi)$ of a general form 
and do not include the derivatives. 
The ERG flow equation reduces to a non-linear second order 
partial differential equation which symbolically can be written as 
\beq
 \dot{f}(\phi,t) = 
{\cal F} (f'' (\phi,t), f' (\phi,t), f(\phi,t);\eta,d), 
\label{ERGE-gen}
\eeq
where $t=-\ln (\Lambda / \Lambda_{0})$ is the flow parameter, 
$\Lambda_{0}$ is a scale where initial conditions are set, 
\[ 
    f(\phi,t) = V_{\Lambda}'(\phi) \equiv 
    \frac{\partial V_{\Lambda}(\phi)}{\partial \phi},
\]
and $\eta$ is the anomalous dimension. 
The prime denotes derivation with respect to the field $\phi$ and 
the dot denotes derivation with respect to the parameter $t$.  
The concrete form of the ERG equation will be given in
Sect.~\ref{sect:N1}. 

FP solutions $f_{*}(\phi)$ satisfy Eq.\ (\ref{ERGE-gen}) with 
$\dot{f}=0$, i.e. the equation 
\beq     
{\cal F} (f'' (\phi), f' (\phi), f(\phi);\eta,d) = 0.  
\label{ERGE-FP}
\eeq   
For the LPA to be consistent the value of the anomalous dimension in the FP 
equation must be set to $\eta_{*}=0$. Eq.\ (\ref{ERGE-FP}) is supplied 
with two boundary conditions, usually at $\phi = 0$. Assuming that the 
theory is symmetric under $Z_{2}$-transformations 
$\phi \rightarrow -\phi$, these conditions are often chosen as 
\beq
f(0) = 0, \; \; \; f'(0) = \gamma.   \label{bc}
\eeq
A consistent ERG FP equation always has the trivial solution 
$f_{*}=0$. It describes the free massless theory and is called the 
Gaussian FP (GFP). For all known formulations the ERG equation is
stiff and most of its solutions singular at some finite value of the field, 
making them not acceptable from the physical point of view. 
Moreover, it turns out that only finitely many solutions of 
Eq.\ (\ref{ERGE-FP}) do not end up in a singularity, thus giving massless 
continuum limits with the prescribed field content \cite{Mo94}. 
They correspond to particular values $\gamma = \gamma_{*}^{(n)}$, 
$n=1,2, \ldots$. 

Let us consider Eq.\ (\ref{ERGE-FP}) in a wider mathematical context 
and study more general solutions, namely solutions for arbitrary $\eta$. 
They are parameterized by $\gamma$ and $\eta$ (for 
$d$ fixed), i.e. $f=f(\phi;\gamma,\eta)$. 
The dependence of ${\cal F}$ on $\eta$ is smooth enough, as a consequence 
$f(\phi;\gamma,\eta)$ depends at least continuously on $\gamma$ and $\eta$. 
It is easy to understand that regular 
solutions correspond to certain points $(\gamma,\eta)$ in the space 
of parameters which form continuous curves $\eta = \eta_{n} (\gamma)$. 
The special values $\gamma_{*}^{(n)}$, which were already introduced above,  
satisfy $\eta_{n} (\gamma_{*}^{(n)})=0$. In what follows, by abuse 
of terminology, we will be referring to these regular solutions of 
Eqs.\ (\ref{ERGE-FP}) and (\ref{bc}) with arbitrary $\eta$ as FP 
solutions, keeping in mind that physical FPs in the LPA 
correspond to $\eta =0$. Such curves of regular solutions 
were first found in Ref. \cite{KNR}.

Obviously, it is useful to know the complete 
space of solutions of the ERG equation in the leading approximation. 
There is also a practical reason for studying FP problem 
(\ref{ERGE-FP}), (\ref{bc}) with $\eta \neq 0$. 
The FP equations to next-to-leading order, i.e. the equations 
obtained with both ${\cal O}(\partial^{0})$ 
and ${\cal O}(\partial^{2})$ terms in $S$ taken into account, form a 
system of coupled nonlinear differential equations. They are stiff 
and solving them require a simultaneous fine 
tuning of $\eta$ and $\gamma$. This makes the direct integration of the 
system a hard task, and one of the ways to solve it is to 
use an iterative numerical 
procedure \cite{BHLM}. In some cases, as for example for $d=3$, one 
can perform such iterations starting with a solution of the first 
order FP equation with $\eta = 0$, i.e. with a physical FP.\footnote{This 
solution is the Wilson-Fischer FP for $d=3$.} However, for $d=2$ this 
is no longer possible. It turns out that due to the nature of the 
problem only periodic or singular solutions, neither of which is a 
physical FP, exist for $\eta = 0$ \cite{Mo95,KNR} (see Sect. 2 
for details). For this reason one needs leading order FP solutions 
with $\eta \neq 0$ as initial solutions to start the iteration procedure.  

In the present article we study solutions of the Polchinski ERG equation 
in $O(N)$ scalar field theories. 
Though much research on the structure of the space of solutions 
has been already done, we feel that some questions remain unanswered.
Namely, in all previous articles the condition of regularity 
was imposed in the course of numerical integration of the equation 
(see Refs.~\cite{Mo94,Mo94a,BHLM} and \cite{CT} for algorithms). 
No "analytical" understanding  
of how this condition works and how it gives rise to critical curves 
$\eta_{n} (\gamma)$ was gained. Also, the relation between the curves and 
the asymptotics of the solutions $f(\phi)$ for large $\phi$ remains obscure. 
To explain this last issue we note that solutions of the  
Polchinski ERG equation in the LPA have, for large $\phi$, 
the asymptotic form \cite{BHLM}
\beq
f(\phi;\gamma,\eta) \sim \left( 1 - \frac{\eta}{2} \right)\phi + 
\tilde{D}(\eta,\gamma)  \phi^{-\frac{\Delta^{-}}{\Delta^{+}}} + \cdots , 
\label{FP-asymp}
\eeq
where we introduced 
\beq
\Delta^{\pm} = 1 \pm \frac{d}{2} - \frac{\eta}{2}. \label{delta-def}
\eeq
The coefficient $\tilde{D}(\eta,\gamma)$ in expansion (\ref{FP-asymp}) 
cannot be derived from the asymptotic analysis of Eq.\ (\ref{ERGE-FP}) alone, 
without solving the FP equation. Fixing $\tilde{D}$ is equivalent 
to setting the boundary conditions at $\phi = \infty$, and only a 
countable number of values of this coefficient correspond to FP 
solutions. The values depend on $\gamma$ and $\eta_{n}$ 
in some intrinsic way.  
In the present article we will address these and other related issues. 

We found it convenient to study them in the context of $O(N)$ scalar 
theories. The reason is that for $N = \infty$ 
a general solution of the lowest order ERG flow equation is 
known analytically though indirectly \cite{CT,Ma1,DKN}. 
Namely, the inverse function $\phi = \phi (f,t)$ can be obtained 
in a closed analytical form. This will be enough for our purpose. We will 
study various properties of the corresponding FP solutions for 
$N = \infty$ and relate them to the analogous properties of FP 
solutions for finite $N$, where only numerical or approximate 
analytical results are available. In particular, we will analyze the 
appearance of the critical curves $\eta_{n} (\gamma)$ of regular 
solutions for $N = \infty$ and see how they change for finite $N$.

We would like to note that $O(N)$ scalar models (also called spherical 
models) are interesting on their own \cite{ZinJu}. They play an important 
role in the description of phase transitions. In particular, $N=4$ 
is relevant for the QCD phase transition \cite{Ma,CVN,Zum94}. 
These models were also 
studied within the ERG using the $1/N$-expansion \cite{RTW} 
(see Refs.\ \cite{BTW} and \cite{MoTu} for other studies and 
references therein).   

The plan of the article is the following. 
In Sect.~\ref{sect:N1} we introduce basic notations and discuss the 
Polchinski ERG equation in the $O(N)$ scalar model for finite $N$. 
Sect.~\ref{sect:Ninf} is devoted to the Polchinski ERG equation in the 
limit $N=\infty$. 
We formulate the condition of regularity of solutions, 
find regular solutions and study their properties in detail. The appearance 
of critical curves $\eta=\eta_{n}(\gamma)$ is analyzed. 
In Sect.~\ref{sect:Nlim} we analyze the behavior of the critical 
curves for finite $N$ and show that they match the $N=\infty$ case. 
Sect.~\ref{sect:concl} contains some discussion and concluding remarks. 
 
\section{$O(N)$ Polchinski ERG equation}

\label{sect:N1}

In this article we consider a general $O(N)$-symmetric scalar 
field theory in $d$-dimen\-sional Euclidean space. 

Let us denote the field components by $\phi^{a}(x)$, $a=1,2, \ldots , N$ 
and assume that the Wilson effective action at the scale $\Lambda$ is 
of the following general form: 
\beq
S_{\Lambda}[\phi] = \frac{1}{2} \int \frac{d^{d}p}{(2\pi)^{d}}   
  \phi_p \cdot  M(p^{2},\Lambda) \cdot \phi_{-p} + 
S_{int,\Lambda}[\phi].\label{wa1}
\eeq
Here $\phi^{a}_{p}$ denotes the field in the momentum 
representation, i.e. the Fourier transform of $\phi^{a}(x)$. In fact the 
field depends on the scale $\Lambda$ as well, i.e. 
$\phi^{a}_{p} = \phi^{a} (p;\Lambda)$. The dots 
in Eq.\ (\ref{wa1}) stand for the contraction over indices of $O(N)$. The 
inverse propagator is
\bea
M_{ab}(p^{2},\Lambda) & = & \delta_{ab} \frac{1}{P(p^2,\Lambda)},\nonumber \\ 
P(p^2,\Lambda) & = & \frac{K({p^2}/{\Lambda^2})}{p^2},  \nonumber 
\eea
where $K(p^2/\Lambda^2)$ is a cutoff function (also called a regulating 
function or simply a regulator) \cite{Po,BaTh}. It has to satisfy  
the following two requirements: 
(1) $K(0)=1$, and (2) $K(z) \rightarrow 0$ as 
$z \rightarrow \infty$ fast enough. The second property guarantees  
that contributions of high momentum modes are suppressed. 
In some studies on the ERG a sharp cutoff function, defined 
by $K(z)=1$ for $0 \leq z \leq 1$ and $K(z)=0$ for $z > 1$, was used 
\cite{WH,HH}. This choice corresponds to integrating out modes 
with $p^{2} > \Lambda^{2}$ exactly. In the present article we assume that 
$K(z)$ is a smooth enough function, so that higher momentum modes 
are integrated out effectively. The  term $S_{int,\Lambda}[\phi]$ is 
the interaction term. In general it contains all possible $O(N)$-invariant 
interactions. 

\subsection{Polchinski ERG equation for arbitrary $N$}

The central idea of the ERG approach is essentially the following: under 
a change of the scale $\Lambda$ the Wilson effective action changes 
in such a way so that the generating functional of the theory and, therefore, 
the Green functions remain unchanged \cite{WH,Po}. This requirement 
gives rise to an equation on $S_{\Lambda}[\phi]$ which is called an ERG flow 
equation. One of the most widely used versions is the Polchinski equation: 

\bea
\Lambda\frac{d}{d\Lambda}S[\varphi;\Lambda] & = & 
\frac{1}{2}\int \frac{d^{d}p}{(2\pi)^{d}} 
\Lambda \frac{d P(p^2,\Lambda)}{d\Lambda} 
\left[\frac{\delta S}{\delta{\varphi_{-p}}}\cdot 
\frac{\delta S}{\delta{\varphi_p}} \right. \nonumber \\
& - & \frac{1}{N}\mbox{Tr}\left(\frac{{\delta^2}S}
{\delta{\varphi_{-p}}\delta{\varphi_p}}\right)  
- \left. 2{P^{-1}}({p^2},\Lambda){\varphi_p}\cdot\frac{\delta S}
{\delta{\varphi_p}}\right],   \label{Peq-1}
\eea 
where we introduced $\varphi_p^{a}=\phi_p^{a}/\sqrt{N}$ and
$S=S[\varphi;\Lambda]=(1/N)S_{\Lambda}[\phi]$, the notations 
which will be convenient in 
the $N \rightarrow \infty$ limit. This equation for $N=1$ was first derived 
and discussed in Ref. \cite{Po}, for $N > 1$ it was studied in Ref. \cite{CT}. 

The structure of Eq.\ (\ref{Peq-1}) becomes more transparent if it 
is rewritten in terms of the dimensionless 
momentum $\hat{p}=p/\Lambda$ and the dimensionless field variable 
$\hat{\varphi}_{p}^{a} = \Lambda^{1+d/2} \varphi_{p}^{a}$. We also introduce 
the renormalization group flow parameter $t = - \ln (\Lambda/\Lambda_{0})$, 
where $\Lambda_{0}$ is some fixed ultraviolet reference scale.
When expanded in powers of $\hat{\varphi}_{p}^{a}$ 
a general action of interaction has the form 

\bea
S_{int}[\hat{\varphi};t] & = & \sum_{k=1}^{\infty} 
\int \frac{d^{d}\hat{p}_{1} \ldots d^{d}\hat{p}_{2k}}{(2\pi)^{d(2k-1)}} 
s_{n;a_{1}, \ldots a_{2k}} (\hat{p}_1,\ldots,\hat{p}_{2k};t) \nonumber \\
 & \times & \hat{\varphi}_{p_1}^{a_{1}}  
\hat{\varphi}_{p_2}^{a_{2}} \ldots \hat{\varphi}_{p_{2k}}^{a_{2k}} 
 \delta^{(d)} (\hat{p}_1 + \ldots + \hat{p}_{2k}).  \label{wa2}
\eea
Note that the terms of the expansion contain only an even number of fields 
and are $O(N)$-invariant. The Polchinski ERG equation, Eq.\ (\ref{Peq-1}), 
becomes 

\bea
\frac{\partial S}{\partial t} & = & \int \frac{d^{d}p}{(2\pi)^{d}} 
K'(\hat{p}^2) \left[\frac{\delta S}{\delta \hat{\varphi}_{-p}} \cdot 
\frac{\delta S}{\delta \hat{\varphi}_p} - \frac{1}{N} \mbox{Tr} 
\left(\frac{{\delta^2}S}{\delta{\hat{\varphi}_{-p}} 
\delta{\hat{\varphi}_p}}\right)\right] \nonumber \\
 & + & S d + 
\int \frac{d^{d}p}{(2\pi)^{d}} \left[1-\frac{d}{2}-\frac{\eta (t)}{2}-
2{\hat{p}^2}\frac{K'({\hat{p}^2})}{K({\hat{p}^2})}\right]
{\hat{\varphi}_p} \cdot \frac{\delta S}{\delta{\hat{\varphi}_p}} \nonumber \\
 & - &  \int \frac{d^{d}p}{(2\pi)^{d}} {\hat{\varphi}_p} \cdot 
\hat{p} \cdot \frac{\partial'}{\partial\hat{p}} 
\frac{\delta S}{\delta{\hat{\varphi}_p}}.  \label{Peq-2}
\eea
The partial derivative in the l.h.s. acts on the 
t-dependent coefficient functions 
$s_{n;a_{1}, \ldots a_{2k}}$ in Eq.\ (\ref{wa2}) and does not act on 
$\hat{\varphi}_{p}^{a}$. The scale dependence of the field 
$\hat{\varphi}_{p}^{a}$ is determined by 

\[
\frac{d \hat{\varphi}_p^{a}}{dt} = 
\left[-1-\frac{d}{2}+\frac{\eta}{2} \right] \hat{\varphi}_p^{a},
\]
where $\eta$ is the anomalous dimension, and has already been taken into 
account in Eq.\ (\ref{Peq-2}) by adding the rescaling terms (the second 
line in Eq.\ (\ref{Peq-2})). The prime in the momentum derivative in the last 
term in the r.h.s. means that it does not act on the $\delta$-function of 
the energy-momentum conservation (see Eq.\ (\ref{wa2})). 

Being supplied with an initial condition 
$S[\hat{\varphi};t=0] = S_{0}[\hat{\varphi}]$ at $t=0$ (or 
$\Lambda = \Lambda_{0}$), where 
$S_{0}[\hat{\varphi}]$ is some given functional, Eq.~(\ref{Peq-2}) 
determines the flow of the effective action and allows to calculate, 
at least in principle, the action and physical characteristics of the theory 
at low energies, when $\Lambda \rightarrow 0$ (or $t \rightarrow +\infty$). 
The limit $\Lambda_{0} \rightarrow \infty$, when the cutoff scale 
$\Lambda_{0}$ is removed, is called the continuum limit and corresponds 
to the renormalized theory. 

\subsection{Leading order approximation}

As was mentioned already in the Introduction, to find solutions 
of the ERG equation in concrete calculations various approximate 
(but non-perturbative) techniques are used. A reliable and efficient 
one is the derivative expansion. In the coordinate representation 
it corresponds to the following expansion of the action  of interaction 
in powers of derivatives: 

\bea
S_{int}[\hat{\varphi};t] & = & 
\int{d^d}\hat{x}
\left[V \left(\hat{\varphi}^{2};t \right) +Z\left(\hat{\varphi}^{2};t \right) 
\left(\frac{\partial \hat{\varphi}(x)}{\partial
    \hat{x}_{\mu}}\right)^{2}
\right. \nonumber \\
 & + & \left. W \left(\hat{\varphi}^{2};t \right)  
\left( \hat{\varphi}(x) \cdot 
\frac{\partial \hat{\varphi}(x)}{\partial \hat{x}_{\mu}} \right)^{2} 
+ \ldots \right], \nonumber
\eea  
where the potentials $V$, $Z$ and $W$ do not contain derivatives of the field. 
We also took into account that these functions depend on the field only 
through the $O(N)$-invariant combination 
$\hat{\varphi}^{2} = \hat{\varphi} (x) \cdot \hat{\varphi} (x) 
\equiv \hat{\varphi}^{a}(x)\hat{\varphi}^{a}(x)$. Restricting the action 
to the first term gives the leading-order of the derivative expansion, 
the LPA.
In this approximation the Polchinski ERG equation reduces to a partial 
differential equation for the function $V(z^2;t)$, where we denoted 
$\hat{\varphi}^{2} = z^{2}$. Introducing the notation
$f(z,t) \equiv \frac{\partial }{\partial z} V(z^2;t)$,
one can write the equation as   

\beq
\dot{f} = \frac{1}{N} f''-2ff'+\left[\left(1-\frac{1}{N}\right)
\frac{1}{z}+{\Delta^{-}}z\right]f'+\left[{\Delta^{+}}-
\left(1-{\frac{1}{N}}\right){\frac{1}{z^2}}\right]f,   \label{P-N}
\eeq 
where the dot and prime denote the derivative with respect to the flow 
parameter $t$ and variable $z$ respectively and, as before, 

\[
\Delta^{\pm} = 1 \pm \frac{d}{2} - \frac{\eta}{2} 
\]
(see Eq.\ (\ref{delta-def})). FP solutions $f(z)$ are solutions of Eq.\ 
(\ref{P-N}) with the property $\dot{f}=0$, i.e. they satisfy the 
equation 

\beq
\frac{1}{N} f''-2ff'+\left[\left(1-\frac{1}{N}\right)
\frac{1}{z}+{\Delta^{-}}z\right]f'+\left[{\Delta^{+}}-
\left(1-{\frac{1}{N}}\right){\frac{1}{z^2}}\right]f = 0.    \label{ONfpe2}
\eeq
Note that $\eta$ is constant here. 

In this paper we will assume that the potential $V(z^{2})$ is analytic 
in $z^{2}$ at $z^{2}=0$, i.e. $V(0)$ is finite and its expansion 
at $z=0$ is convergent and contains only non-negative integer powers of 
$z^{2} = \hat{\phi}^{2}$. This assumption is fulfilled 
in most of scalar models and, therefore, does not add new essential 
restrictions in field theory applications of the ERG. Correspondingly, 
the expansion of $f(z)$ at $z=0$ includes only odd powers of $z$. 
To have a unique FP solution we must add two conditions on $f(z)$. In 
field theory problems it is natural to impose them at $z=0$ in the form 

\beq
f(z){|_{z=0}}=0,\qquad f'(z){|_{z=0}}=\gamma.     \label{ONic1}
\eeq
For purely historical reasons these conditions are called initial conditions. 
The first condition is a consequence of the $O(N)$-symmetry of the 
potential $V(\hat{\varphi}^{2})$. The parameter $\gamma$ plays the 
role of the square of the mass of the field. 

For further studies it is convenient to define a new variable $y$ and a new 
function $u(y,t)$ instead of $z$ and $f(z,t)$ in the following way: 

\bea
z & = & \sqrt{2y},   \label{x-y} \\
f(z,t) & = & \sqrt{\frac{y}{2}} u(y,t).   \label{f-u}
\eea
Note that these formulas define $u(y,t)$ on the semi-axes $y > 0$ only.
In terms of $y$ and $u(y,t)$ Eq.\ (\ref{P-N}) becomes  

\beq
\dot{u}(y,t) = \frac{2y}{N} u_{yy}'' - 2yuu_{y}' - u^{2} + 
\left( 1 + \frac{2}{N} + 2y\Delta^{-} \right) u_{y}' + su, \label{P-N-1} 
\eeq
where we introduced the notation 

\beq
s \equiv \Delta^{+} + \Delta^{-} = 2 - \eta.  \label{s-def}
\eeq

The prime in Eq.\ (\ref{P-N-1}) denotes the derivative with respect to
$y$. The FP equation now reads 

\beq 
\frac{2y}{N} u_{yy}'' - 2yuu_{y}' - u^{2} + 
\left( 1 + \frac{2}{N} + 2y\Delta^{-} \right) u_{y}' + su = 0.  \label{P1f}
\eeq
Let us rewrite initial conditions (\ref{ONic1}) in terms of $u(y)$. 
Using relations (\ref{x-y}), (\ref{f-u}) it is easy to check that 
the condition $f(0)=0$ together with the assumption about 
the behavior of $f(z)$ at $z=0$ are equivalent to the requirement 
that $u(y)$ is analytic at $y=0$, i.e. $u(0)$ is finite and its 
expansion at $y=0$ is convergent and contains only non-negative 
integer powers of $y$. Therefore, from now on we will be interested 
only in solutions of Eq.\ (\ref{P1f}) which are analytic at $y=0$. 
The second initial condition in (\ref{ONic1}) 
transforms to 

\beq
u(y){|_{y=0}}=2\gamma.     \label{ONic2}
\eeq
Using the analyticity of $u(y)$ the value of $u'(y)$ at the origin 
can be calculated from the FP equation (\ref{P1f}): 

\[
u'(y){|_{y=0}}=\frac{2\gamma (2\gamma-s)}{1+\frac{2}{N}}.
\]

In fact, to be physical solutions functions $f(z)$ or $u(y)$ must 
satisfy even stronger requirements. 
Recall that $f(z)$ is the derivative of the effective potential 
$V$ with respect to the field variable. In field theory applications one is 
interested in solutions such that $V(\hat{\varphi}^{2})$ and its derivatives 
are finite and do not have singularities for all finite values of the 
field variable. This means that $f(z)$ does not have singularities 
for finite $z$ or, equivalently, $u(y)$ does not have singularities for 
finite $y \geq 0$. Such solutions will be called {\it regular} solutions. 

{}From Eqs.\ (\ref{ONfpe2}) and (\ref{ONic1}), or, equivalently, from 
Eqs.\ (\ref{P1f}) and (\ref{ONic2}), one can see that for a given $d$ a 
FP solution is completely determined by the values of $\gamma$ and $\eta$. 
Correspondingly, any solution of the FP problem can be 
represented by a point $(\gamma,\eta)$ in the parameter space. 

One comment is in order. Since the potentials $Z$ and $W$ are not 
taken into account in the LPA, the kinetic term in the effective action,
which contains two derivatives of the field, matches ERG equation 
(\ref{Peq-2}) for $\eta = 0$ only. Regular solutions of 
Eqs.\ (\ref{ONfpe2}), (\ref{P1f}) with $\eta=0$ will be called 
{\it physical} FPs. As we have already explained in the 
Introduction, we will relax 
this condition and study a wider class of FP solutions with $\eta \neq 0$. 
To summarize, in the present article we will study solutions 
$u(y)$ of Eq.\ (\ref{P1f}) with arbitrary $\eta$, both regular and 
non-regular, which satisfy condition (\ref{ONic2}) and are analytic 
at $y=0$. 

The asymptotics of a solution $u(y)$ for large $y$ can be 
obtained from ERG equation (\ref{P1f}) by a straightforward calculation. 
Namely, one writes a general expression for the expansion of $u(y)$ as a  
series in decreasing powers of $y$ and fixes the coefficients and the 
exponents from Eq.\ (\ref{P1f}). One gets 

\beq
u(y) = s - D y^{\beta} 
 - \frac{\alpha D^{2}}{s(2-\alpha)} y^{2\beta} - \frac{D^{3}}{s^{2}} 
\frac{\alpha (\alpha + 1)}{(2-\alpha)^{2}} y^{3\beta}  
- D \frac{1 + \frac{2}{N}\beta}{s(2-\alpha)^{2}} y^{\beta -1}+ \cdots, 
 \label{uy-as}
\eeq
where 

\[
\alpha = \frac{2 \Delta^{-}}{s} = 
\frac{2 \Delta^{-}}{\Delta^{+}+\Delta^{-}}, \; \; \; 
\beta = - \frac{1}{2-\alpha} = - \frac{s}{2 \Delta^{+}}.
\]
As was already mentioned in the Introduction, 
this technique does not allow 
to fix the constant $D=D(\gamma,\eta)$. Expansion (\ref{uy-as}) is, 
of course, in a full correspondence with asymptotic formula 
(\ref{FP-asymp}) with 

\[
D = 2^{-\frac{1}{2} \left( \frac{\Delta^{-}}{\Delta^{+}}-1 \right)} 
\tilde{D}.
\]

The FP equation has two trivial solutions. The first one is obtained for 
$\gamma=0$ and is equal to $f=0$. 
This is the GFP describing the free 
massless theory. It corresponds to the line $\gamma=0$ on the 
$(\gamma,\eta)$-plane. The second solution is obtained for $\gamma = s/2$ and 
is given by the linear function $f(z)= sz/2$, 
where $s$ is defined in Eq.\ (\ref{s-def}). These solutions,  
termed high temperature FPs or infinite mass GFPs, are described 
by the straight line $\eta = 2 - 2\gamma$ on the 
$(\gamma,\eta)$-plane. For the sake of brevity we will call them trivial  
FPs (TFP). As one can easily see from Eqs.\ (\ref{f-u}) and
(\ref{x-y}), these solutions correspond to $u=0$ and $u=s$ respectively. 

\subsection{$N=1$ case} 

Before starting the analysis of FP solutions in the $O(N)$ model for  
arbitrary $N$ it is instructive to consider the $N=1$ model.
In this case the $O(N)$-symmetry reduces to the $Z_{2}$-symmetry under  
the reflection $\phi \rightarrow -\phi$. The leading order Polchinski 
ERG equation (\ref{ONfpe2}) with initial condition (\ref{ONic1}) 
becomes the familiar FP problem (see \cite{BHLM})

\bea
f''(z)-2f(z)f'(z)&+&{\Delta^-}zf'(z) + {\Delta^+}f(z) = 0,
\label{Z2fpe1} \\
f(0)=0,& & \; \; \; f'(0)=\gamma. \label{ONic3}
\eea

For $\Delta^{-}=0$ or $\Delta^{+}=0$ this equation can be integrated 
analytically. Let us analyze the case $\Delta^{-}=0$ first. With this 
condition, which implies $\Delta^{+}=d$, Eqs.\ (\ref{Z2fpe1}) and
(\ref{ONic3}) become

\beq
f''(z)-2f(z)f'(z)+df(z) = 0, \; \; f(0)=0, \; \;  f'(0)=\gamma.  \label{Z2fpe2} 
\eeq 
Denoting $g=f'$ one can rewrite Eq.\ (\ref{Z2fpe2}) as a first order 
differential equation for $g(f)$:

\beq 
g \frac{dg}{df} = f \left[ 2g(f) - d \right], \quad g(f)|_{f=0} =
\gamma. 
\label{g-f} 
\eeq
If $\gamma < d/2$ solutions to problem (\ref{g-f}) are given by the 
following periodic trajectories in the phase space $(f,f')=(f,g)$:

\[
f^{2}=g-\gamma+\frac{d}{2}\ln\left(\frac{{d}/2-g}{{d}/2-\gamma}\right).
\]
In particular, for $|\gamma| \ll d$ the trajectories are given
approximately by ellipses 

\[
d \cdot f^{2} + g^{2} = \gamma^{2}. 
\]
This family of regular solutions corresponds to the straight line 
$\eta = 2 - d$ for all $\gamma < d/2$ on the $(\gamma,\eta)$-plane. 
Note that for $d=2$ they are solutions with $\eta=0$. 

If $\gamma \geq d/2$ the trajectories in the phase space are unbounded. 
They describe solutions $f(z)$ which are singular at some finite value 
of $z$. 

Let us consider now the case $\Delta^{+}=0$, i.e. $\eta = 2+d$. 
Eqs.\ (\ref{Z2fpe1}) and (\ref{ONic3}) become 

\beq
f''(z) - 2f(z)f'(z) - dzf'(z) = 0, \; \; f(0)=0, \; \;  f'(0)=\gamma.
\label{Z2fpe3} 
\eeq
Solutions of this problem can be obtained from solutions of Eq.\ 
(\ref{Z2fpe2}) with $\gamma$ substituted by $\gamma + d/2$. Indeed, 
let function $h(z)$ be a solution of 

\[
h''(z) - 2h(z)h'(z) + d h(z)=0, \;\; h(0)=0, \; \; h'(0)=\gamma+\frac{d}{2} 
\]
(cf. (\ref{Z2fpe1})). As one can easily check, then $f(z)=h(z)- (d/2) z$ 
is a solution of problem (\ref{Z2fpe3}). 

In a general case solutions of (\ref{Z2fpe1}), (\ref{ONic3}) can
be found only numerically. A generic solution $f(z)$ or its derivative 
end up with a singularity at a finite point $z=z_{0}$, i.e., is 
not regular. 
A natural method for searching regular solutions is to fine tune $\eta$ 
and $\gamma$ in such a way that the position of singularity 
$z_{0}=z_{0}(\gamma,\eta) \rightarrow \infty$. 
A realization of this recipe in a practical computation is to pick a 
solution for which $z_{0}$ shows a tendency to diverge \cite{Mo94,BHLM}. 

First we studied numerically solutions of the FP problem, 
Eqs.\ (\ref{Z2fpe1}) and (\ref{ONic3}), for $d=2$, in a wide range of values 
of $\gamma$, $\eta$. We found that for $\eta=0$ and any $\gamma < 1$ 
there is a periodic solution. This is in accordance with our analytic result 
for $\Delta^{-}=0$ above (note that for $d=2$ $\Delta^{-}=-\eta/2$). 

\begin{figure}[ht]
\psfig{file=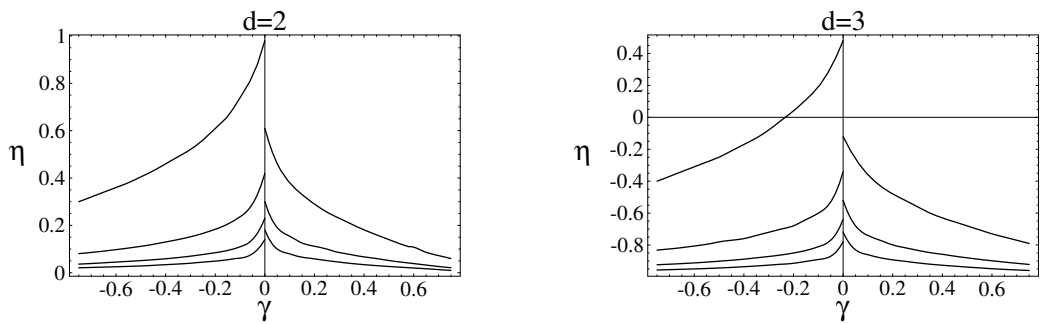,width=\hsize}
%\epsfxsize=1.0\hsize
%\epsfbox{plotd32a.eps}
\vspace{-1cm}
\caption{Non-trivial FP curves for $d=2$ and $d=3$. 
Only the first 7 curves are shown.}
\label{fig:d32a}
\end{figure}

For $0 < \eta \leq 1$ and $\eta < 2 - 2\gamma$ the parameter $\gamma$ can 
be fine tuned to a certain value so that a non-trivial regular FP 
solution exists. The set of such values $(\gamma,\eta)$ form an infinite 
discrete series of curves $\eta_{n}(\gamma)$ ($n=1,2,3, \ldots$) 
shown in Fig.~\ref{fig:d32a}. At $\gamma=0$ $\eta_{n}(0)=2/(n+1)$. 
The curves accumulate at the line $\eta=0$ as $n \rightarrow \infty$. 

There are general grounds to expect the existence of the curves 
$\eta_{n}(\gamma)$ whose points $(\gamma,\eta_{n}(\gamma))$ correspond 
to regular FP solutions \cite{KNR}. Indeed, let us suppose that for some 
value $(\gamma',\eta')$ there exists a regular solution. This means that 
the singularity of this solution is situated at $z_{0}(\gamma',\eta') = 
\infty$. Consider now another value $\eta''$ sufficiently close to 
$\eta'$. Assuming that the function $z_{0}(\gamma,\eta)$ is continuous in the 
vicinity of $(\gamma',\eta')$ it is clear that there exists the value 
$\gamma''$ such that again $z_{0}(\gamma'',\eta'') = \infty$. Consequently, 
there is a curve $\eta(\gamma)$ passing through the point $(\gamma',\eta')$ 
in the parameter space. The curve is defined by the equation 
$z_{0}(\gamma,\eta) = \infty$. 

We found that when moving along a given curve $\eta_{n}(\gamma)$ 
the shape of the corresponding solution qualitatively remains the same. When passing 
from one curve to another the shape of the solution changes significantly 
following the same regular pattern as the one obtained by T.R.~Morris 
within the Legendre ERG equation \cite{Mo95}. Since all the curves 
$\eta_{n}(\gamma)$ are situated in the $\eta > 0$ half-plane, their 
presence is a signal of the existence of an infinite number of 
non-trivial FPs in two-dimensional scalar theories. This 
expectation turns out to be true. As it was shown in Refs.\
\cite{Mo95} and \cite{KNR} by studying the next-to-leading 
ERG equations, they do correspond to the minimal unitary series of 
$p(p+1)$ ($p=3,4, \ldots$) conformal models. 

For $d=3$ we found periodic solutions corresponding to $\eta=-1$. 
Again, this is in accordance with our analytic results for 
$\Delta^{-}=0$ because for $d=3$ the condition $\Delta^{-}=0$ leads to 
$\eta=-1$. We also found an infinite number of critical curves 
$\eta=\eta_{n}(\gamma)$, $n=1,2,\ldots$, corresponding to 
non-trivial regular FP solutions of Eqs.\ (\ref{Z2fpe1}) and (\ref{ONic3}). 
They are plotted in Fig.~\ref{fig:d32a}. The curves 
accumulate at the line $\eta=-1$ as $n \rightarrow \infty$. At $\gamma=0$ 
$\eta_{n}(0)=(2-n)/(n+1)$. The curve $\eta_{1}(\gamma)$ crosses 
the $\gamma$-axis at $\gamma_{*}=-0.229 \ldots$. The value of $\gamma$ 
coincides with the one for which a non-trivial 
physical FP was found in the LPA of the Polchinski equation 
\cite{BHLM}. This is the well-known Wilson-Fischer FP analyzed in 
numerous previous studies (see \cite{HH},\cite{Mo94}-\cite{HKLM},
\cite{Co98}, \cite{Feld}-\cite{TeWe94}. It 
belongs to the Ising model universality class. The rest 
of the curves are situated between the horizontal lines 
$\eta=0$ and $\eta=-1$ (except for the value of $\eta_{2}(\gamma)$ 
at $\gamma=0$: $\eta_{2}(0)=0$, the GFP). This suggests that for 
$d=3$ there is only one non-trivial physical FP, namely the 
one associated with the curve $\eta_{1}(\gamma)$. 

\section{$N=\infty$ Polchinski ERG equation}

\label{sect:Ninf}

In this section we will study solutions of the leading order Polchinski 
ERG equation for $N = \infty$. In this limit many features of the 
solutions can be described analytically, and we are going to take 
advantage of this. Our aim is to study the $\eta (\gamma)$ curves of 
regular FP solutions for $N = \infty$. We expect them to be certain 
smooth deformations of the analogous curves for finite $N$, therefore   
the analysis of the $N = \infty$ case will help us to understand 
the features of the $\eta (\gamma)$ curves for finite $N$. 

\subsection{Equation and general solution \label{sec:eq_gen_sol}}

In the limit $N \rightarrow \infty$ the Polchinski flow equation,  
Eq.\ (\ref{P-N-1}), simplifies considerably and becomes  

\beq
\dot{u}(y,t) = - 2yuu_{y}' - u^{2} + 
\left( 1 + 2y\Delta^{-} \right) u_{y}' + su \label{P-inf} 
\eeq
for $-\infty < t < \infty$ and $y \geq 0$ (see Eq.\ (\ref{x-y})).  

The FP equation for $u(y)$ follows from Eqs.~(\ref{P1f}) and (\ref{ONic2}),
and is given by 

\bea
- 2yuu_{y}' - u^{2} + 
\left( 1 + 2y\Delta^{-} \right) u_{y}' + su &=& 0,  \label{P1} \\
u(y)|_{y=0} &=& 2\gamma.    \label{P1-ic}
\eea
As in the case of finite $N$,
Eq.\ (\ref{P1}) is solved by the GFP
$u(y)=0$ and by the TFP $u(y)=s$.

Let us consider solutions of flow equation (\ref{P-inf}) first. 
As was pointed out in Ref.~\cite{CT}, it can be 
solved analytically for the inverse function $y(u,t)$. Indeed, 
from Eq.\ (\ref{P-inf}) one obtains the following 
flow equation for $y(u,t)$: 

\beq
\dot{y}(u,t) = (u-s) u y_{u}' - 2 \left( \Delta^{-} - u \right) y -1. 
\label{yt-eq}
\eeq
Its general solution is equal to \cite{CT}

\beq
y(u,t) = \frac{1}{(s-u)^{2-\alpha} u^{\alpha}}  
F\left( \frac{s-u}{u}e^{st} \right) + y(u),  \label{yut}
\eeq
where 

\beq
\alpha=2{\Delta^-}/s, \label{def-alpha}   
\eeq
$F$ is an arbitrary function and $y(u)$ satisfies the inverse FP equation, 
i.e. Eq.\ (\ref{yt-eq}) with $\dot{y}=0$, and the corresponding 
initial condition:   

\bea
(u-s)u{{y'}_u}-2\left({\Delta^-}-u\right)y(u)-1 &=& 0, \label{ifpe} \\
y(2\gamma) &=& 0.    \label{coiy}   
\eea
The FP equation, of course, can also be obtained directly by inverting 
Eq.\ (\ref{P1}). Condition (\ref{coiy}) follows 
from (\ref{P1-ic}). 

Eq.\ (\ref{ifpe}) can be solved analytically \cite{CT}. 
Its general solution is given by

\beq
y(u) = \frac{1}{(s-u)^{2-\alpha} u^{\alpha}} \left[ 
- \frac{u^{\alpha} (s-u)^{1-\alpha}}{\alpha} 
+ \frac{\alpha - 1}{\alpha} {\int_0^u} 
\left( \frac{s - z}{z} \right)^{-\alpha} dz  - C \right]. \label{yu1}
\eeq 
Note that it is consistent with general solution (\ref{yut}) 
with the function $F$ taken to be constant, $F=-C$. 

The form of the solution is not unique and integrating by parts one 
can bring it to another one. Possible divergences of the integral in 
Eq.\ (\ref{yu1}) can be absorbed in the integration constant $C$. 
Equivalently, one can choose the lower limit to be $u_{0} \neq 0$, 
in this case $C$ becomes $u_{0}$-dependent. 

It is important to bear in mind that a solution $u(y)$ may not be 
monotonous and, therefore, not invertible on the whole semi-axis 
$y \geq 0$. As an example let us consider the case in which 
$u(y)$ is monotonically 
decreasing in a subinterval $0 \leq y \leq y_{\#}$ and monotonically 
increasing in the infinite interval $y_{\#} \leq y < \infty $. One has 
$u(0) > u(y_{\#})$ and $u(\infty) > u(y_{\#})$.  In each of these regions 
$u(y)$ is invertible, let us denote the corresponding inverse functions 
as $y_{(<)}(u)$ and $y_{(>)}(u)$ respectively. The former is defined for 
$u_{\#} \leq u \leq u(0)$, where $u_{\#} \equiv u(y_{\#})$, 
the latter for $u_{\#} \leq u < u(\infty)$.  
They have to satisfy the following matching condition:    

\beq
y_{(<)}(u_{\#}) = y_{(>)}(u_{\#}).  \label{mc}
\eeq
Hence, there are two branches of $y(u)$, $y_{(<)}(u)$ and $y_{(>)}(u)$, 
such that their inverse functions form the solution $u(y)$ for 
$0 \leq y < \infty$. 
Note that initial conditions (\ref{P1-ic}) and (\ref{coiy}) imply that 
$y_{(<)}(u(0)) = y_{(<)}(2\gamma)=0$. This equality fixes the 
integration constant $C$ in formula (\ref{yu1}) for the $y_{(<)}(u)$ 
branch. The analogous constant in the expression for the $y_{(>)}(u)$ 
branch is fixed by the matching condition, Eq.\ (\ref{mc}). 
Concrete examples will be considered in Sect.\ \ref{sec:C_ne_0_al_1over_n}.

To conclude this discussion we stress once more 
that Eq.\ (\ref{yu1}) gives the exact general solution of the Polchinski 
FP equation, Eq.\ (\ref{P1}), in 
the LPA in terms of the inverse function $y(u)$. 
The constant $C$ is fixed either by initial condition (\ref{coiy}) or by 
a matching condition similar to (\ref{mc}). 

The parameter $\alpha$, defined by (\ref{def-alpha}), turns out 
to be very convenient in the analysis of FP solutions. The 
anomalous dimension $\eta$ and the parameter $s$ are expressed 
in terms of $\alpha$ as follows: 

\beq
\eta = 2 - \frac{d}{1-\alpha}, \; \; \; 
s = \frac{d}{1-\alpha}.    \label{eta-s}
\eeq
There are different types of solutions for different intervals of 
values of $\alpha$. The interval which is most interesting for physical 
applications is 

\beq
-1 < \alpha < 0.      \label{alpha-range}
\eeq
It corresponds to the range $2 - d < \eta < (4-d)/2$. From now on 
we will only be considering interval (\ref{alpha-range}) of 
values of $\alpha$. The form of the solution, given by Eq.\ (\ref{yu1}), 
is adequate to this interval.  

The asymptotics of the solution $u(y)$ for large $y$ is given by the 
$N \rightarrow \infty$ limit of Eq.\ (\ref{uy-as}). In the case in hand 
it can also be found from exact formula (\ref{yu1}). 
For this we first calculate the asymptotic behavior of 
the function $y(u)$ as $u \rightarrow s$ and then invert the asymptotic 
formula. This procedure not only reproduces formula (\ref{uy-as}), 
but also allows to determine the constant $D$. One obtains 

\[
D = \left[ \frac{s^{\alpha}}
{C - \frac{\pi d}{\sin \pi \alpha}} \right]^{-\frac{s}{2\Delta^{+}}}.  
\]

Now we turn to the analysis of the expression for the derivative of the 
solution $u(y)$ following from Eq.\ (\ref{P1}):

\beq
 u_{y}' = \frac{u(s-u)}{2uy - (1+2y\Delta^{-})}.   \label{uder-u} 
\eeq
The denominator is zero if $2yu = (1+2y\Delta^{-})$. Let us introduce the 
function
\beq
u_{sing}(y) = \frac{1+2y\Delta^{-}}{2y}.   \label{using-def}
\eeq
For $\alpha$ satisfying (\ref{alpha-range}) we have $s>0$ and $\Delta^- < 0$.
It follows that a general solution $u(y)$ with initial condition
$u(0)=2\gamma$ (see (\ref{P1-ic})) 
and asymptotics $\lim_{y\to\infty}u(y)=s$ (from Eq.\ (\ref{uy-as}))
necessarily crosses the curve of potential 
singularities $u_{sing}(y)$ at some point $y=y_{sing}$.
In other words,
there always exists at least one point $y_{sing} > 0$ such that 
$u(y_{sing}) = u_{sing}(y_{sing})$. Let us denote this value of $u$ as 
$u_{0}$,

\[
u_{0} \equiv u(y_{sing}) = \frac{1+2y_{sing}\Delta^{-}}{2y_{sing}}. 
\]
Since in general $u_{0} \neq 0$, 
$u_{0} \neq s$ the derivative of the solution does not exist at this 
point and the solution is singular.\footnote{This observation was 
made in Ref.~\cite{CT}.}

Note that this picture has an equivalent 
description in terms of the inverse function $y(u)$. 
Its derivative is obtained from Eq.\ (\ref{ifpe}),

\[
{y'_u}=\frac{1+2({\Delta^-}-u)y}{u(u-s)}.
\]
So the line of potential singularities is 

\beq
{y_{sing}}(u)=\frac{1}{2(u-{\Delta^-})},\label{ysing-def}
\eeq
the inverse function of ${u_{sing}}(y)$. 

\begin{figure}[tp]
\center{
\psfig{file=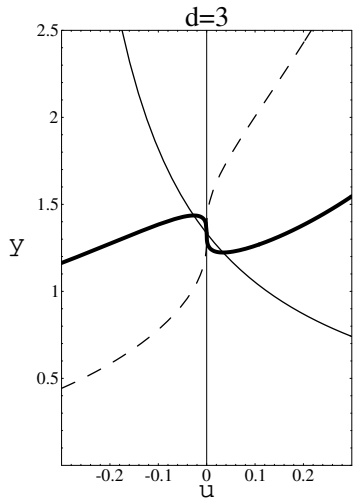,width=1.0\hsize}}
%\epsfxsize=0.9\hsize
%\epsfbox{pony12si.eps}
\caption{Plot of the curve of potential singularities 
$y_{sing}(u)$ defined by Eq.\ (\ref{ysing-def}) for $d=3$ and 
${\Delta^-}=-3/8$. Also given, respectively as bold and dashed curves, are the plot of a singular solution $y(u)$ satisfying the initial condition $y(2\gamma)=0, \gamma=-5/2$ and the plot of a regular solution $y(u)$ satisfying 
$y(2\gamma^{*})=0, {\gamma^*}=-1/2$. The point where all these curves intersect is $(0,{y_*}=-1/(\alpha s))$ with $\alpha=-1/3$ and $s=9/4$.}
\label{fig:using}
\end{figure} 
A general solution $y(u)$ always intersects this curve at 
some point, namely at $u=u_{0}$ (see Fig.~\ref{fig:using}). 
Note that in general there may be a few points of intersection. 
If ${u_0}\not=0, {u_0}\not= s$ the derivative 
${y'_u}$ vanishes, the derivative ${u'_y}$ does not exist, and the 
solution is singular.

The analysis of relation (\ref{uder-u}) shows that a general solution 
has the following behavior near the point of singularity:

\beq
u(y) = u_{0} +  A \sqrt{y-y_{sing}} + b (y-y_{sing}) + c (y-y_{sing})^{3/2} 
+ \cdots. 
\label{uy-sing} 
\eeq 
The coefficients are determined from Eq.\ (\ref{uder-u}) and are equal to 

\[
A = \pm \sqrt{\frac{u_{0}(s-u_{0})}{y_{sing}}}, \; \; \; 
b = \frac{s+\Delta^{-} - 3u_{0}}{3y_{sing}}, \; \; \mbox{etc.}  
\]
Using these expressions one can easily see that the solution is indeed 
non-analytic and the derivative does not exist at $y=y_{sing}$: 

\beq
u'(y) = \frac{A}{2\sqrt{y-y_{sing}}} + b + \cdots . \label{u1-sing}
\eeq
By inverting a general solution, 
Eq.\ (\ref{uy-sing}), one gets the following expansion for $y(u)$ in the 
vicinity of $u=u_{0}$: 

\[ 
y(u) = y_{sing} + \frac{1}{A^{2}} (u - u_{0})^{2} - 
\frac{2b}{A^{4}} (u-u_{0})^{3} + 
\left( \frac{5b^{2}}{A^{6}} - \frac{2c}{A^{5}} \right) (u-u_{0})^{4} + 
\cdots. 
\]
The corresponding derivative is 

\[
{y'_u}=\frac{2}{A^2}(u-{u_0})-\frac{6B}{A^4}{(u-{u_0})^2}+
4\left( \frac{5b^{2}}{A^{6}} - \frac{2c}{A^{5}} \right) (u-u_{0})^{3} + 
\cdots, 
\]
which at $u={u_0}$ is indeed zero.

The situation changes if the parameter $\gamma$ in the initial condition,
Eq.\ (\ref{P1-ic}), is adjusted to a value $\gamma^{*}$ such 
that $u(y)$ crosses the curve of 
potential singularities at $y=y_{*}$, where $y_{*}$ is determined by the 
condition $u(y_{*}) = u_{sing}(y_{*}) = 0$. From Eq.\ (\ref{using-def}) 
one can see that 

\beq
y_{*} = -{1\over{\alpha s}}. \label{y*}
\eeq
Such a solution $u(y;\gamma^{*})$ does not have root 
behavior and $u'(y)$ is regular at the point of potential singularity 
$y=y_{*}$. Consequently ${y'_u}$ is no longer zero at $u=0$ 
(see Fig.~\ref{fig:using} where a plot of $y(u;\gamma^{*})$ satisfying 
$y(0;\gamma^{*})=y_{*}$ is also given).   

The expansion of a regular solution in the vicinity of $y=y_{*}$ 
contains only positive integer powers of $(y-y_{*})$ with the leading 
term being 

\beq
u(y) = b_{n} (y - y_{*})^{n} + \; \mbox{higher powers of $(y-y_{*})$}. 
\label{u-reg}
\eeq
It is the leading power $n$ which defines the type of the solution. 
It turns out that there are two qualitatively different classes of 
regular solutions: (1) with $n=1$, and (2) with $n \geq 2$. 

For $n=1$ the expansion at $y=y_{*}$ is given by 

\bea
u(y) & = & b_{1} (y-y_{*}) + b_{2} (y-y_{*})^{2} + b_{3}(y-y_{*})^{3} 
+ \cdots,  \label{u-reg1} \\
b_{1} & = & - \frac{s^{2} \alpha (\alpha+1)}{2}, \; \; \; 
b_{2} = - \frac{3 s^{3} \alpha^{2} (\alpha + 1)^{2}}{4 (\alpha +2)}.  
\label{b1b2}
\eea
For $n \geq 2$, as one can find out using ERG FP equation (\ref{P1}) or 
Eq.\ (\ref{uder-u}), the behavior of the solution is more 
complicated: 

\bea
u(y) & = & b_{n} (y-y_{*})^{n} + c_{2n} (y-y_{*})^{2n-1} + 
b_{2n} (y-y_{*})^{2n} \label{u-reg2}  \\
 & + & d_{3n} (y-y_{*})^{3n-2} + c_{3n} (y-y_{*})^{3n-1} + \cdots .
 \nonumber 
\eea
For such a function to satisfy the equation the parameter $\alpha$ 
must be fixed to the value $\alpha = - 1/n$. 
This amounts to $\eta = 2 - dn /(n+1)$, $s=dn/(n+1)$, etc. However, 
the coefficient $b_{n}$ is not fixed by the equation and remains a free 
parameter.  In other words, there is a regular solution for any 
value of $b_{n}$. The rest of the coefficients are expressed in terms of 
$b_{n}$. For example,  

\[
c_{2n} = - \frac{2n^{3}}{s^{2}(n-1)} b_{n}^{2}, \; \; \; 
b_{2n} = - \frac{2n+1}{s} b_{n}^{2}, \; \; \; \mbox{etc.}
\]
Note that for $n=2$ the terms proportional to $(y-y_{*})^{2n}$ and to 
$(y-y_{*})^{3n-2}$ give the same power of $(y-y_{*})^{4}$ so 
that expansion (\ref{u-reg2}) is of the form 

\[
u(y) = b_{2} (y-y_{*})^{2} + c_{4} (y-y_{*})^{3} + 
(b_{4}+d_{6}) (y-y_{*})^{4} + \cdots .
\]

Let us analyze the expansion of the corresponding inverse functions 
$y(u)$ around $y_*$.
Consider the two classes of regular solutions. 

\begin{itemize}

\item[(i)] For the solution $u(y)$ with $n=1$ the inverse 
function $y(u)$ has the following behavior close to $u=0$:

\beq
y(u) = y_{*} + \frac{1}{b_{1}} u - \frac{b_{2}}{b_{1}^{3}} u^{2} + 
{\cal O}(u^{3}),     \label{y-reg1}  
\eeq
where $b_{1}$, $b_{2}$ are given by Eq.\ (\ref{b1b2}). 

\item[(ii)]  For regular solutions with $n \geq 2$, $y_{*}=n/s$ and 
the expansion for $y(u)$ is the sum of two series in powers on $u$ 
of the form

\beq
y(u) = \frac{n}{s}+ a u^{\frac{1}{n}} 
\left[ 1 + k_{1} u + k_{2} u^{2} + \ldots \right] + 
l_{1} u + l_{2} u^{2} + \cdots ,  \label{y-reg2}  
\eeq
where $a$, $k_{i}$, $l_{j}$ are related to the coefficients in 
Eq.\ (\ref{u-reg2}) as follows: 

\[
a= \frac{1}{b_{n}^{1/n}}, \; \; \; 
k_{1} = - \frac{b_{2n}}{n b_{n}^{2}} = \frac{2n+1}{sn},  \; \; \; 
l_{1} = - \frac{c_{2n}}{n b_{n}^{2}} = \frac{2n^{2}}{s^{2}(n-1)},  
\; \; \; \mbox{etc.}
\]

\end{itemize}

This analysis explains the nature of the condition of regularity of solutions 
of the ERG equation (for $N=\infty$): the parameter $\gamma$ must be 
adjusted to the value for which the solution $u(y)$, fixed by the initial 
condition $u(0)=2\gamma$, satisfies

\beq
u \left( -\frac{1}{2\Delta^{-}} \right) = 0.  \label{reg-cond}
\eeq
This property actually follows from Eq.\ (\ref{ifpe}). Indeed, it 
is easy to see that if the derivative $y'(u)$ is finite at $u=0$, 
then $y(0)$ must be 

\[
y(0) = -\frac{1}{2 \Delta^{-}} = -{1\over{\alpha s}} = y_{*}.   
\]
The latter is equivalent to (\ref{reg-cond}). 

We would like to note that there is another class of solutions, 
namely those which satisfy 

\[
u(y'_{*}) = u_{sing}(y'_{*}) = s,
\]
where $y'_{*} = 1/(2\Delta^{+})$. As it can be seen from (\ref{uder-u}) and 
(\ref{using-def}), its derivative does not have a singularity and the solution 
is analytic at the point of potential singularity. This class will not be 
discussed further in the article.  

{}From now on we focus on regular solutions of Eqs.\ (\ref{P1}) and 
(\ref{P1-ic}) only. When it is necessary to indicate their dependence 
on the parameters we will be using the extended notation $u(y;\gamma,\eta,d)$ 
and $y(u;\gamma,\eta,d)$ correspondingly. 

Expanding formula (\ref{yu1}) in powers of $u$ around $u=0$ one gets 

\beq
y(u) = -\frac{1}{\alpha s} - \frac{C}{s^{2-\alpha}} \frac{1}{u^{\alpha}} 
\left[ 1 - \frac{\alpha -2 }{s} u + {\cal O}(u^{2}) \right] - 
\frac{2}{\alpha (1+\alpha) s^{2}} u + {\cal O}(u^{2}).   
\label{yu-exp}
\eeq
Compare this expansion with Eqs.\ (\ref{y-reg1}) and (\ref{y-reg2}). 
It is easy to see that there are two possibilities for function 
(\ref{yu-exp}) to give a regular FP solution: 

\begin{itemize}

\item[(i)] $C=0$, in which case solution (\ref{y-reg1}) or (\ref{u-reg1}) 
is reproduced. 

\item[(ii)] $C \neq 0$, $\alpha = - 1/n$, $n=2,3, \ldots$, in which case 
solution (\ref{y-reg2}) or (\ref{u-reg2}) is obtained. 

\end{itemize}

One may also wonder whether there is a regular solution in the case 
$C \neq 0$, $\alpha = -1$. It can be readily checked 
that the answer is negative. 
Indeed, for $\alpha = -1$ the integral in (\ref{yu1}) can be calculated 
explicitly. To avoid the divergence at the lower limit let us consider 
a modification of Eq.\ (\ref{yu1}) with the lower limit of integration 
$u=0$ substituted by $u=u_{0} > 0$ and the constant $C$ 
changed to another constant $C_{u_{0}}$. Performing the calculation 
one gets 

\beq
y(u) = - \frac{u}{(s_{1}-u)^{3}} C_{u_{0}} + \frac{1}{s_{1}-u} - 
\frac{2u(u-u_{0})}{(s_{1}-u)^{3}} + 
\frac{2s_{1}u}{(s_{1}-u)^{3}} \ln \frac{u}{u_{0}},   \label{yu-1}
\eeq
where $s_{1}=d/2$. For $\alpha=-1$ it follows from Eq.\ (\ref{eta-s}) that 
$\eta = (4-d)/2$. The behavior of solution (\ref{yu-1}) at $u=0$ is 
given by 

\[
y(u) = \frac{1}{s_{1}} + \frac{2}{s_{1}^{2}} u \ln u + 
{\cal O} (u,u^{2}\ln u, \ldots ).
\] 
Consequently though $u(1/{s_1})=u'(1/{s_1})=0$ the second 
derivative $u''(y)$ is singular at $y=1/s_{1}$ and so the 
solution $u(y)$ for $\alpha=-1$ is not regular. 

In the following subsections we will see that the two classes of functions 
listed above are quite different and characterized by two distinct sets
of eigenvalues and eigenoperators for perturbations around the FP 
\cite{CT}. Let us consider both cases in turn. 

\subsection{$C=0$} 

In this case the solution $u(y)$ is invertible for all 
$0 \leq y < \infty$. The integration constant in Eq.\ (\ref{yu1}) 
is fixed by initial condition (\ref{coiy}) and, therefore, 
for a given $d$ becomes the following function of $\gamma$ and $\eta$: 

\beq
C(\gamma,\eta) =
- {{{{(2\gamma)}^\alpha}{{(s-2\gamma)}^{1-\alpha}}}
\over{\alpha}}+{{\alpha-1}\over{\alpha}}{\int_0^{2\gamma}}
{{\left({{s-z}\over{z}}\right)}^{-\alpha}} dz.     \label{icons}
\eeq

The condition $C(\gamma,\eta)=0$ defines a curve in the 
parameter space which we denote as $\eta_{1}(\gamma)$. The lower index 
"1" corresponds to the power $n=1$ of the leading term in expansion 
(\ref{u-reg}). Using Eq.\ (\ref{icons}) the condition 
$C(\gamma,\eta)=0$ can be written as 

\beq
-\frac{2\gamma}{d} (1-\alpha)^{2} 
\left(1 - \frac{2\gamma}{d} (1-\alpha) \right)^{\alpha -1} \int_{0}^{1} dz 
z^{\alpha} \left( 1- \frac{2\gamma}{d} (1-\alpha) z \right)^{-\alpha} = 1,  
\label{C-0}
\eeq
where we took into account the second relation in (\ref{eta-s}). 
Solutions in the case under consideration will be denoted as 
$u_{1}(y;\gamma,d) \equiv u(y;\gamma,\eta_{1}(\gamma),d)$ and 
$y_{1}(u;\gamma,d) \equiv y(u;\gamma,\eta_{1}(\gamma),d)$. They 
exist only for negative values of the parameter $\gamma$. 
This can be seen from the following arguments. Eqs.\ (\ref{eta-s}), 
(\ref{alpha-range}) and (\ref{y*}) imply that $s>0$ and $y_{*}>0$. 
In accordance with asymptotic formula (\ref{uy-as}) the 
solution $u_{1}(y;\gamma,d)$ approaches $s$ for large $y$. 
As it follows from Eqs.\ (\ref{u-reg1}) and (\ref{b1b2}), 
$u_{1}(y_{*};\gamma,d)=0$ and has a positive linear slope at $y=y_{*}$. 
Therefore, regular solutions exist for $\gamma < 0$ only, since otherwise 
there would be other zeros of $u_{1}(y)$ in the interval $0 \leq y < y_{*}$ 
and $u_{1}(y_{*};\gamma,d)$ would not be invertible. 

Thus, Eq.\ (\ref{C-0}) defines $\alpha$ as a function of $\gamma$ for 
$\gamma < 0$. In fact, one can easily see that it is a function 
of the ratio $\gamma / d$. We denote it by $\alpha_{1}(\gamma/d)$. 
In accordance with the first relation in Eq.\ (\ref{eta-s}) this  
defines the function 

\beq
\eta_{1} (\gamma) = 2 - \frac{d}{1 - 
\alpha_{1}\left( \frac{\gamma}{d}\right)}. 
\label{eta-h}
\eeq
Limiting values of $\alpha_{1}(\gamma /d)$ 
can be readily derived from Eq.\ (\ref{C-0}). 
For $\gamma < 0$ and $\gamma \rightarrow {0^-}$

\[
\alpha_{1} \left(\frac{\gamma}{d}\right) \rightarrow -1, \; \; \mbox{or} 
\; \; \eta_{1}(\gamma) \rightarrow \frac{4-d}{2}. 
\]
These limiting points correspond to the GFP. Expanding Eq.\ (\ref{C-0}) 
in powers of $\gamma$ and using Eq.\ (\ref{eta-h}) we obtain that 

\beq
\eta_{1} (\gamma) = \frac{4-d}{2}  + 2\gamma + 24 \frac{\gamma^{2}}{d} + 
480 \frac{\gamma^{3}}{d^{2}} + 
d \cdot {\cal O}\left(\frac{\gamma^{4}}{d^{4}}\right). \label{eta1-0}
\eeq
For $\gamma \rightarrow -\infty$ 

\[
\alpha_{1} \left(\frac{\gamma}{d}\right) \rightarrow 0, \; \; 
\mbox{or} \; \; 
\eta_{1}(\gamma) \rightarrow 2-d. 
\]
The asymptotic expansion of the function 
$\alpha_{1} (\gamma /d)$ can be found from the analysis of Eq.\ (\ref{C-0}).
Introducing the variable $w = - d/(2\gamma)$, which is positive and 
tends to zero as $\gamma \rightarrow -\infty$, we get

\[
\alpha_{1} \left(\frac{\gamma}{d}\right) 
= -w + w^{2} - w^{2} \ln w - w^{3} \ln^{2} w +\cdots . 
\]
The dots stand for the terms $w^{3}$, $w^{4}$, $w^{4}\ln ^{3} w$, etc., 
which in this limit are at least logarithmically smaller 
than the terms written down explicitly. 

As before, we denote by $\gamma_{*}$ the point at which 
$\eta_{1}(\gamma) =0$. As was explained in the Introduction, 
this point is of special interest because 
it corresponds to the physical FP solution of the Polchinski 
ERG equation in the LPA. From expansion (\ref{eta1-0}) one 
sees that $\gamma_{*}=0$
for $d=4$. Hence, in this case only the trivial GFP exists. For $d=2$
we find $\gamma_{*} = -\infty$ and, therefore, 
there are no non-trivial physical FP solutions for finite 
values of $\gamma$.

Let us analyze the case $d=3$. The value $\eta =0$ 
corresponds to $\alpha = -1/2$ (see Eq.\ (\ref{eta-h})). The 
integral in formula (\ref{C-0}) can be calculated 
explicitly. One gets the equation 

\beq
\frac{3}{2} \sqrt{2\kappa_{*}} (2 - 2\kappa_{*})^{-3/2} 
\left( \sinh (2\theta_{*}) + 2\theta_{*} \right) = 1,   \label{eta-3}
\eeq
where the parameter $\theta_{*}$ is defined by the relation 

\[
\sinh \theta_{*} = \sqrt{\frac{\kappa_{*}}{2}},     
\]
and $\kappa_{*} = e^{-i \pi} 2\gamma_{*} > 0$. 

\begin{figure}[ht]
\center{
\psfig{file=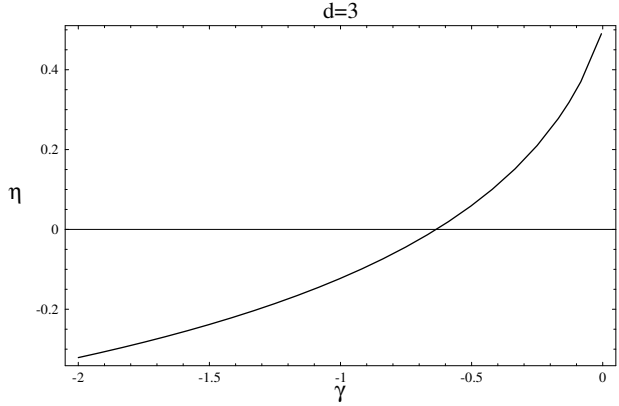,width=0.8\hsize}}
%\epsfxsize=0.9\hsize
%\epsfbox{ploneg3.eps}
\caption{Plot of the function $\eta_{1} (\gamma)$ for $d=3$.}
\label{fig:eta-gamma}
\end{figure} 

\begin{figure}[tp]
\psfig{file=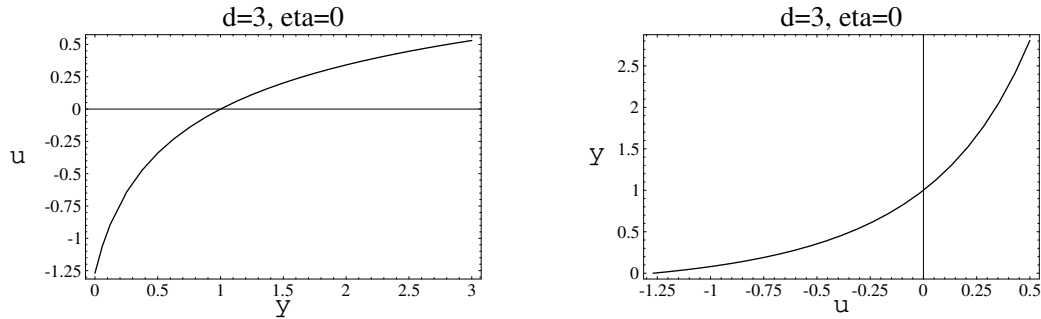,width=1.0\hsize}
\vspace{-1.5cm}
\caption{Plots of the functions $u_1 (y)$ and $y_1 (u)$ for $d=3$ 
and $\eta(\gamma_*)=0$, $\gamma_*=- 0.634913 \cdots$.}
\label{fig:uy1}
\end{figure}

To obtain the curve $\eta = \eta_{1}(\gamma)$ we 
solved Eq.\ (\ref{C-0}) numerically for $\gamma < 0$. 
For $d=3$ the result is given in
Fig.~\ref{fig:eta-gamma}. In other dimensions the function
$\eta_{1}(\gamma)$ has a similar profile and can be easily computed
from the $d=3$ curve using the scaling properties which will be 
obtained in Sect.~\ref{sect:Nlim}. 
The physical FP solution $u_{*}(y) \equiv u_{1}(y;\gamma_{*},3)$ 
corresponds to $\gamma_{*} = - 0.634913 \ldots$ which is the value 
satisfying Eq.\ (\ref{eta-3}). The function $u_{*}(y)$ and its inverse 
$y_{*}(u)$ are given in Fig.~\ref{fig:uy1}. The numerical results 
for the curve ${\eta_1}(\gamma)$ confirm that
non-trivial regular solutions exist for $\gamma<0$ and $-1<\eta<1/2$ 
only. For $\gamma=0$, $\eta=1/2$ we have the GFP. Recall
that the same is true for $N=1$. Note as well that the curves 
${\eta_1}(\gamma)$ for $N=1$ and $N=\infty$ have very similar shape 
(see Ref.~\cite{KNR} and Sect.~\ref{sect:N1}). Since the
curve ${\eta_1}(\gamma)$ has a part with positive values $\eta\geq 0$ 
it corresponds to a physical FP. 

The eigenoperators of perturbations around the FP solution 
$u = u_{1}(y;\gamma,3)$ are equal to 

\beq
{g_m}(y)=\exp\left({\int_0^y}dz \,{t_m}(z)\right), \label{C1-eop}
\eeq
where 

\bea
{t_m} & = & {{{\lambda_m}+2y{{u'}_y}-s+2u}\over{1+2{\Delta^-}y-2yu}}, 
\label{C1-tm} \\
{\lambda_m}& = &-s(m+\alpha).  \label{C1-ev}
\eea
The spectrum (\ref{C1-ev}) and the linear behavior of $u_{*}(y)$ 
at $y=y_{*}$, Eqs.\ (\ref{u-reg1}) and (\ref{b1b2}) indicate that 
the physical FP at $\gamma = \gamma_{*}$ is the Wilson-Fischer 
FP \cite{CT}.

\subsection{$C \neq 0$, $\alpha = - 1/n$ \label{sec:C_ne_0_al_1over_n}}

Now we consider regular FP solutions with $C \neq 0$ and 

\[
 \alpha = \alpha_{n} = -\frac{1}{n}, \; \; \;  n = 2,3,\cdots
\]
in Eq.\ (\ref{yu-exp}). From relations (\ref{eta-s}) we get that in this case 

\bea
s & = & s_{n} =\frac{dn}{n+1},    \nonumber \\
\eta & = & \eta_{n} = 2 - \frac{d}{1-\alpha_{n}} = 
2 - \frac{dn}{n+1}.  \label{eta-n}
\eea

In what follows we will use the notations $u_{n}(y;\gamma,d)$ and 
$y_{n}(u;\gamma,d)$ 
for the FP solutions $u(y;\gamma,\eta_{n},d)$ and $y(u;\gamma,\eta_{n},d)$ 
respectively. It follows from Eqs.\ (\ref{u-reg2}) and (\ref{yu-exp}) 
that the leading behavior of the solutions in the vicinity of $u=0$ 
is given by 

\[
y_{n}(u;\gamma,d) = y_{*n} + \frac{C}{s_{n}^{2+1/n}} u^{n} + \cdots ,
\] 
and in the vicinity of the point of potential singularity $y=y_{*n}=n/s_{n}=
n(n+1)/d$ is equal to 

\[
u_{n}(y;\gamma,d) = \frac{s_{n}^{-(2n+1)}}{C^{n}} (y - y_{*n})^{n}  
+ \cdots .    \label{uy-b}
\] 
The curves of regular solutions are straight lines 
$\eta = \eta_{n}=$ const in the $(\gamma,\eta)$-plane. We consider the 
cases of even $n$ and odd $n$ separately. 

\subsubsection{$n$ even}

In this case a solution $y_{n}(u;\gamma,d)$ has two branches. As it was 
discussed earlier, the branch which satisfies 
$0 \leq y_{n}(u;\gamma,d) \leq y_{*n}$ is given by formula (\ref{yu1}) 
with the integration constant equal to 

\[
C_{n}(\gamma) = - \frac{(2\gamma)^{\alpha_{n}} (s_{n}-2\gamma)^{1-\alpha_{n}}}
{\alpha_{n}} + \frac{\alpha_{n}-1}{\alpha_{n}} \int_{0}^{2\gamma}
{{\left({{s_{n}-z}\over{z}}\right)}^{-\alpha_{n}}} dz     
\]
(cf. (\ref{icons})). This expression is a consequence of the initial 
condition, Eq.\ (\ref{coiy}). 

\begin{figure}[tp]
\center{
\psfig{file=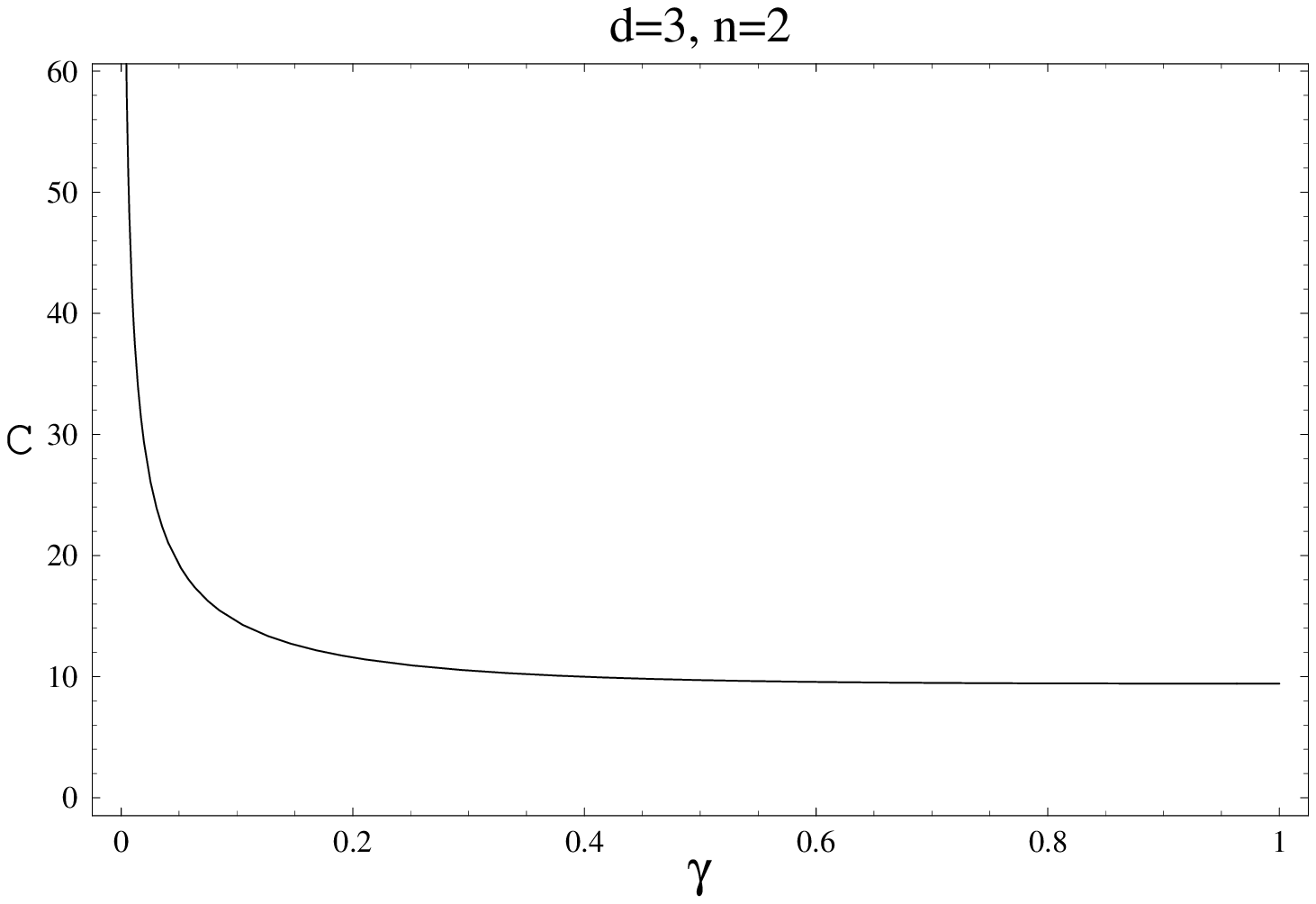,width=0.8\hsize}}
%\epsfxsize=0.9\hsize
%\epsfbox{poncgp32.eps}
\caption{Plot of the function $C_{2} (\gamma)$ for $d=3$.}
\label{fig:C2}
\end{figure}

For $n$ even the function $C_{n}(\gamma)$ and the solution $y_{n}(u;\gamma,d)$ 
exist for $0 < \gamma < s_{n}/2$. In particular, for $n=2$ one gets 

\[
C_{2} (\gamma) = + 3s_{2} \arcsin \sqrt{\frac{2\gamma}{s_{2}}} +  
2 \sqrt{\frac{s_{2} - 2\gamma}{2\gamma}} (s_{2} + \gamma),  
\]
with $s_{2}=2d/3$. The plot of the function $C_{2}(\gamma)$ for $d=3$ is 
presented in Fig.~\ref{fig:C2}. 

\begin{figure}[tp]
\psfig{file=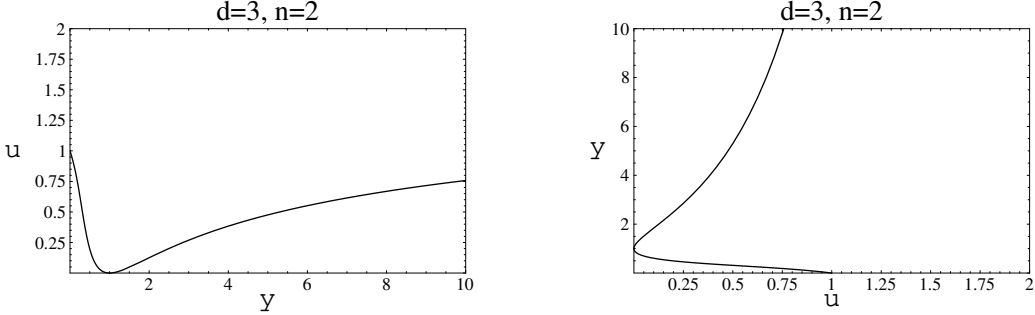,width=1.0\hsize}
%\epsfxsize=0.9\hsize
%\epsfbox{pon32uy.eps}
\vspace{-1.5cm}
\caption{Plots of the functions $u_2 (y)$ and $y_2 (u)$ for $d=3$ 
and $\gamma = 1/2$.}
\label{fig:uy2}
\end{figure}

For each value of $\gamma \geq 0$ and $\gamma < s_{n}/2$ the two branches 
of the solution $y_{n}(u;\gamma,d)$ are given by: 

for $0 \leq u \leq 2\gamma$, $0 \leq  y_{n}(u) \leq y_{*n}$ 

\bea
y(u) & = & - \frac{C_{n}(\gamma)}{u^{\alpha} (s-u)^{2-\alpha}} 
- \frac{1}{\alpha (s-u)} + \frac{\alpha-1}{\alpha} 
\frac{1}{u^{\alpha} (s-u)^{2-\alpha}} 
\int_{0}^{u} \left(\frac{s-z}{z}\right)^{-\alpha} dz \label{yn-even1} \\
& = & - \frac{u^{1/n}}{(s-u)^{2+1/n}} C_{n}(\gamma)
- \frac{n}{(s_{n}-u)} + \frac{(n+1) u^{1/n}}{(s_{n}-u)^{2+1/n}} 
\int_{0}^{u} \left(\frac{s_{n}-z}{z}\right)^{-\alpha} dz; \nonumber 
\eea

for $0 \leq u \leq s$, $y_{*n} \leq y_{n}(u) < \infty$

\bea
y(u) & = & \frac{C_{n}(\gamma)}{u^{\alpha} (s-u)^{2-\alpha}} 
- \frac{1}{\alpha (s-u)} + \frac{\alpha-1}{\alpha} 
\frac{1}{u^{\alpha} (s-u)^{2-\alpha}} 
\int_{0}^{u} \left(\frac{s-z}{z}\right)^{-\alpha} dz \label{yn-even2} \\
& = & \frac{u^{1/n}}{(s-u)^{2+1/n}} C_{n}(\gamma)
- \frac{n}{(s_{n}-u)} + \frac{(n+1) u^{1/n}}{(s_{n}-u)^{2+1/n}} 
\int_{0}^{u} \left(\frac{s_{n}-z}{z}\right)^{-\alpha} dz. \nonumber 
\eea
Formulas (\ref{yn-even1}), (\ref{yn-even2}) match together at $u=0$. 
These two branches define a unique function $u_{n}(y;\gamma,d)$ for 
$y \geq 0$. 
In Fig.~\ref{fig:uy2} we show the plots of the
functions ${u_2}(y;1/2,3)$ and ${y_2}(u;1/2,3)$ respectively. 

It is easy to see that $C_{n}(\gamma) \rightarrow +\infty$ when 
$\gamma \rightarrow {0^+}$. This limit, of course, corresponds  
to the GFP. For $\gamma = s_{n}/2$ we find 

\[
C_{n}(s_{n}/2) =  nd \frac{\pi \alpha_{n}}{\sin (\pi \alpha_{n})} = 
 \frac{\pi d}{\sin \frac{\pi}{n} }. 
\]
The corresponding regular solution is $u_{n}(y;s_{n}/2,d)=s_{n}=$ const, 
the TFP. This function is not invertible, hence the corresponding solution 
$y_{n}(u;s_{n}/2,d)$ does not exist at $\gamma = s_{n}/2$. 

\subsubsection{$n$ odd}

\begin{figure}[tp]
\center{
\psfig{file=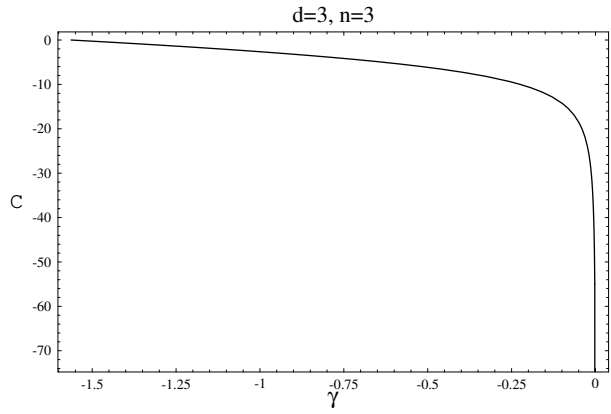,width=0.8\hsize}}
\caption{Plot of the function $C_{3} (\gamma)$ for $d=3$.}
\label{fig:C3}
\end{figure}

\begin{figure}[tp]
\psfig{file=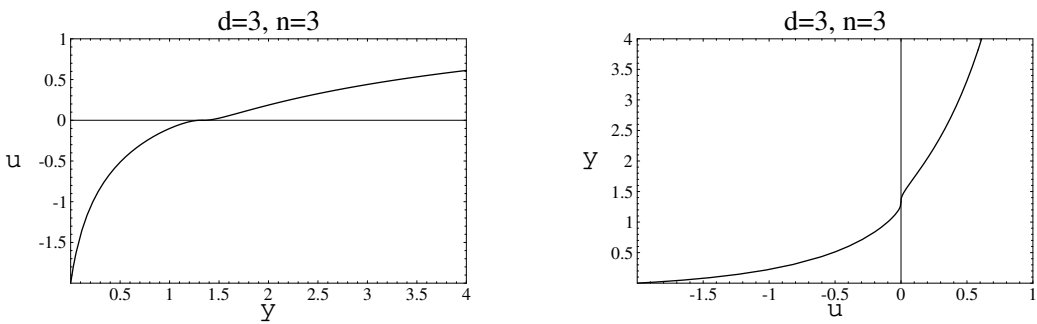,width=1.0\hsize}
\vspace{-1.5cm}
\caption{Plots of the functions $u_{3} (y)$ and $y_3 (u)$ for 
$d=3$ and $\gamma = -1$.}
\label{fig:uy3}
\end{figure}

For $n$ odd, $n \geq 3$, regular solutions exist only for negative 
$\gamma$ and consist of one branch. The function $y_{n}(u;\gamma,d)$ is 
defined for $2\gamma \leq u < s$ and varies in the range 
$0 \leq y_{n}(u;\gamma,d) < \infty$. The integration constant $C$ in 
formula (\ref{yu1}) is fixed by the initial condition, Eq.(\ref{coiy}), 
and is given by Eq.\ (\ref{icons}). As before, we denote it by 
$C_{n}(\gamma)$. It is easy to see that for odd $n$ 
as  $\kappa = e^{-i\pi n} 2\gamma \rightarrow {0^+}$ 

\[
C_{n}(\gamma) \sim -n \kappa^{-\frac{1}{n}} s_{n}^{1-\alpha_{n}} 
\rightarrow -\infty,
\]
The plot of the function $C_{3}(\gamma)$ for $d=3$ is shown in 
Fig.~\ref{fig:C3}. The solutions $u_{3}(y;-1,3)$ and $y_{3}(u;-1,3)$ 
are given in Fig.~\ref{fig:uy3}.  

Note the characteristic cubic root behavior of $y_{3}(u;-1,3)$ in 
the vicinity of $u=0$ (see Eq.\ (\ref{yu-exp})). 

At certain value $\gamma = \gamma_{*n}<0$ the function $C_{n}(\gamma)=0$. 
Of course, this is the point where the straight line $\eta = \eta_{n}$  
($n=3,5, \ldots$) in the $(\gamma,\eta)$-plane hits the curve 
$\eta = \eta_{1}(\gamma)$ obtained in Sect.\ \ref{sec:eq_gen_sol}. 

\begin{figure}[tp]
\center{
\psfig{file=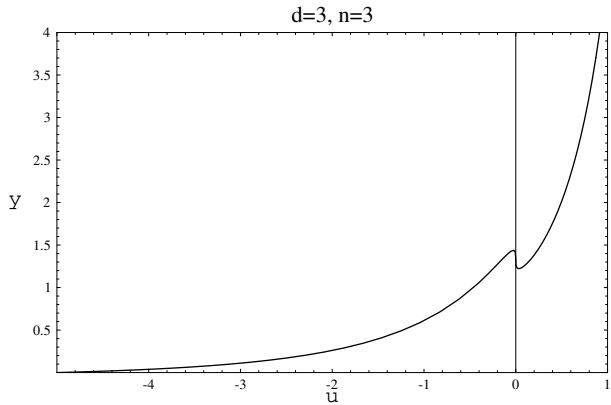,width=0.8\hsize}}
\caption{Plot of the function $y_{3} (u)$ for $d=3$ and 
$\gamma=-2.5$. The function is non-monotonous in a vicinity of $u=0$.}
\label{fig:yu3d}
\end{figure}

The solution 
$u_{n}(y;\gamma_{*n},d)$ coincides with $u_{1}(y;\gamma_{*n},d)$ for 
this value of $\gamma$. When trying to search for solutions 
for $\gamma < \gamma_{*n}$ one finds out that $y_{n}(u;\gamma,d)$ becomes 
non-monotonous, and hence non-invertible, in a vicinity of $u=0$, and 
therefore the solution $u_{n}(y;\gamma,d)$ does not exist. An example 
of such behavior is illustrated in Fig.~\ref{fig:yu3d} which shows the plot 
of the function $y_{3}(u;-2.5,3)$. 

The eigenoperators are of the same form as in Eqs.\ (\ref{C1-eop}) and
(\ref{C1-tm}). The eigenvalues are given by 

\beq
{\lambda^n_m}={s_n}(1-\frac{m}{n}).
\eeq
Note that these are also the eigenvalues of the GFP.   
\vspace{0.5cm}

Let us summarize the results obtained in this section. 
We have found the infinite discrete series of lines $\eta = \eta_{n}(\gamma)$ 
($n=1,2,3,\ldots $) located in a certain region of the parameter space 
$(\gamma,\eta)$. Points on these lines correspond to regular FP solutions 
of the $N=\infty$ Polchinski ERG equation in leading approximation. The 
curve $\eta_{1}(\gamma)$ (see Fig.~\ref{fig:eta-gamma}) bounds the region 
from the left. From the right the region is bounded by the straight line 
$\eta = 2-2\gamma$ of TFP solutions $u=s$. The curves $\eta=\eta_{n}$, 
$n=2,3, \ldots$, are straight horizontal lines (see Eq.\ (\ref{eta-n})). They 
start at $\gamma=0$ (GFP solution) and end up either at 
$\gamma = s_{n}/2 > 0$ (the TFP solution) for $n$ even, or at 
$\gamma = \gamma_{*n} < 0$ for $n$ odd. 
The value $\gamma_{*n}$ is defined by the condition 
$\eta_{1}(\gamma_{*n}) = \eta_{n}$. As $n \rightarrow \infty$ the lines 
$\eta = \eta_{n}$ accumulate at $\eta_{\infty}=2-d$. 

\section{Solutions for finite $N$ and their $N\rightarrow\infty$ 
limit}

\label{sect:Nlim}

In the sections above we have seen a detailed treatment of the
Polchinski ERG equation for $N=1$ and $N=\infty$. Not presented
there are the finite $N$ cases, that are of interest in their own
right, but also because the $N=\infty$ case should be viewed as a
limiting case for large $N$. For that reason, we present in this
section a study of the finite $N$ case, and show how the $N=\infty$
case comes about as a limit.

First of all from the plots for $N=1$ in Fig.~\ref{fig:d32a} it is 
easy to notice that the patterns of the curves of regular solutions are 
the same for $d=2$ and $d=3$. They differ just by a vertical shift. 
Our analysis in the previous section shows that this is also true in the 
$N=\infty$ case. It turns out that for a given $N$ the pattern of the curves 
$\eta_{n}(\gamma)$ is universal and does not depend on the dimension of the 
space $d$. Namely, when we pass from one number of dimensions $d$ to another 
$\tilde{d}$ the functions $\eta_{n}(\gamma)$ experience a constant shift and 
some scaling transformation while the pattern of the curves is preserved. 
This is a consequence of a property of the ERG equation which we are 
going to analyze in the remainder of this section.

Let $u(y;\gamma,\eta,d)$ be a regular solution of (\ref{P1f}), (\ref{ONic2}) 
corresponding to a point of a curve $\eta = \eta_{n}(\gamma)$. 
Perform a scaling transformation 

\[
u(y) \rightarrow \tilde{u}(\tilde{y}) = \lambda u(y;\eta,\gamma,d), 
\; \; \; \tilde{y} = \frac{1}{\lambda} y. 
\]
It is easy to check that the function $\tilde{u}(\tilde{y})$ satisfies 
the equation  

\bea
\frac{2\tilde{y}}{N} \frac{d^{2}}{d\tilde{y}^{2}} \tilde{u}(\tilde{y}) 
- 2\tilde{y}\tilde{u} \frac{d \tilde{u}}{d\tilde{y}} 
&-&\tilde{u}^{2} + 
\left( 1 + \frac{2}{N} + 2\tilde{y}\Delta^{-} \lambda \right) 
\frac{d \tilde{u}}{d\tilde{y}} + s \lambda \tilde{u} = 0,  \label{P1f-scale} \\
&&\tilde{u}(0)=2\tilde{\gamma} \equiv 2 \lambda \gamma.     
\label{ONic2-scale}
\eea
Therefore $\tilde{u}(\tilde{y})$ is a {\it regular} solution in 
$\tilde{d}$ dimensions corresponding to the point 
$(\tilde{\gamma},\tilde{\eta})$ of the curve 
$\tilde{\eta}=\tilde{\eta}_{n}(\tilde{\gamma})$, where $\tilde{\gamma}$, 
$\tilde{\eta}$, $\tilde{d}$ and a relation between $\eta_{n}(\gamma)$ and 
$\tilde{\eta}_{n}(\tilde{\gamma})$ can be found from 
Eqs.\ (\ref{P1f-scale}) and (\ref{ONic2-scale}) as follows. According to Eqs.\ 
(\ref{delta-def}) and (\ref{s-def}), $\eta=2 - s$, $d=s - 2\Delta^{-}$. 
Correspondingly, 
$\tilde{\eta}=2 - \tilde{s}$, $\tilde{d}=\tilde{s} - 2\tilde{\Delta}^{-}$, 
where $\tilde{\Delta}^{-}= \lambda \Delta^{-}$, $\tilde{s}=\lambda s$, 
see Eq.\ (\ref{P1f-scale}). Combining these relations we obtain that 

\bea
\tilde{d} & = & \lambda d, \; \; \; \tilde{\gamma}= \lambda \gamma, 
\nonumber \\
\tilde{\eta} & = & 2 - \lambda s = 2 - \lambda (2 -\eta) =  
 2 (1 - \lambda) + \lambda \eta. \nonumber 
\eea
Trading $\lambda$ for $\tilde{d}/d$ we arrive at the relation 

\beq
\tilde{\eta}_{n}(\gamma) = 2 \left( 1 - \frac{\tilde{d}}{d} \right) + 
\frac{\tilde{d}}{d} \eta_{n} \left(\frac{d}{\tilde{d}} \gamma \right). 
\label{eta-eta}
\eeq
This formula is valid for any $N \geq 1$. 

Eq.\ (\ref{eta-eta}) can be rewritten as the relation 

\[
\frac{\tilde{\eta}_{n} (\tilde{d}\gamma) -2}{\tilde{d}} = 
\frac{\eta_{n} (d\gamma) -2}{d}
\]
which tells that the combination 
$\beta_{n}(\gamma) \equiv (\eta_{n}(d\gamma)-2)/d$ 
is in fact independent of $d$. Therefore, a general solution of 
functional equation (\ref{eta-eta}) is given by 

\beq
\eta_{n}(\gamma) = 2 + d \beta_{n} \left( \frac{\gamma}{d} \right). 
\label{eta-sol}
\eeq

The functions $\eta_{n}(\gamma)$ studied in the previous 
sections do have this property. 
We checked that in the case $N=1$ and $d=2$ or $d=3$, the curves 
$\eta_{n} (\gamma)$ verify relation (\ref{eta-eta}) or, equivalently, 
are described by (\ref{eta-sol}). Special limits $\Delta^{+}=0$, 
$\Delta^{-}=0$ are consistent with (\ref{eta-eta}), (\ref{eta-sol}) 
as well.
The results for the $N \rightarrow \infty$ case in Sect. 3 are also 
in a full correspondence with this property. In particular, from 
Eqs.\ (\ref{eta-h}) and (\ref{eta-n}) it follows that 

\[
\beta_{n} \left( \frac{\gamma}{d} \right) = 
- \frac{1}{1- \alpha_{n}\left( \frac{\gamma}{d} \right)}. 
\]
Recall that for $n \geq 2$ $\alpha_{n}$ is the constant function 
$\alpha_{n}=-1/n$. 

For $d$ fixed the shape of the curves changes with $N$. For $N=1$, $d=2,3$,   
a few curves are shown in Fig.~\ref{fig:d32a}. For $N=\infty$, $d=3$  
the plot of $\eta_{1}(\gamma)$ is given in Fig.~\ref{fig:eta-gamma}.  
For $n \geq 2$ $\eta_{n}=2 - dn/(n+1)=\hbox{const}$, i.e., they are 
represented by horizontal lines. 
It is of interest to study the curves for finite $N > 1$ and understand 
the transition from $N=1$ to $N=\infty$. 
In particular, a natural question arises: Do the intersections of 
the curves $\eta_n(\gamma)$
($n$ odd, $n \geq 3$) with the curve $\eta_1(\gamma)$, encountered for
$N=\infty$, have an equivalent for finite $N$? 
Such intersections do not exist for $N=1$. 
Is there a value of $N$ for which
intersections are encountered, or the $N=\infty$ curves
approach the large $N$ limit in some other way? 

To gain more insight into the properties of the functions 
$\eta_{n}(\gamma)$ for an arbitrary $N$ it is useful to study their 
behavior in the vicinity of $\gamma=0$. This can be done by a 
perturbative analysis for small values of $\gamma$. 
To this end, perform the change of variables 
$v=-N \Delta^{-}y$, $u(y)=2\gamma h(v)$. Eqs.~(\ref{P1f}) and 
(\ref{ONic2}) turn into

\bea
L(v)h-E\,h&+&{2\gamma\over\Delta^-}v\,h\,h'_z+{\gamma\over\Delta^-}h^2 = 0,
\label{eq:sys1}\\
&& h(0)=1,        \label{eq:sys2}
\eea
where

\[
E={s\over2\Delta^-}, \; \; L(v)h=v\,h^{''}_{vv}+(M-z)h'_v  
\]
and 

\[
M={N\over2}+1.
\]

The linearized equation $L(v)h=E\,h$ can be solved using the known
solutions of the eigenvalue problem $L(v)h_n=\lambda_nh_n$:

\beq
\lambda_n=-n,\quad h_n(v)=\Phi(-n,M;v),\quad n=0,1,2,\cdots,
\eeq
where

\bea
\Phi(a,b;v)&=&1+{a\over b}v+{(a)_2\over(b)_2}{v^2\over2!}+\cdots,
\nonumber\\
(a)_n&=&a(a+1)\ldots(a+n-1),\qquad (a)_0=1,   \nonumber
\eea
denotes the confluent hypergeometric function (see Ref. \cite{GrRy}). Note that $h_n(v)$ 
is a polynomial of order $n$ with $h_n(0)=1$.

We now develop perturbation theory around these solutions. Fixing $n$,
we write

\bea
h(v)&=&(1+\gamma\,{\cal D}_1+\gamma^2\,{\cal D}_2+\cdots)
(h_n(v)+\gamma\sum_{k\ne n}c_kh_k+\gamma^2\sum_{k\ne n}d_kh_k+\cdots), 
\nonumber \\
E&=&E_n^{(0)}+\gamma E_n^{(1)}+\gamma^2E_n^{(2)}+\cdots\nonumber 
\eea
(here ${\cal D}_n$ are constants)
and solve system (\ref{eq:sys1}), (\ref{eq:sys2}) order by order in
$\gamma$. Note that, at any order, the sums over $k\ne n$ are over a
finite number of terms. Here we list a few results. 

At order $\gamma^0$, we find $E_n=E_n^{(0)}=-n$, from which it follows
that

\beq
\eta_n=2-{n\,d\over n+1}+{\cal O}(\gamma),    \label{eta-0}
\eeq
a result that is independent of $N$.
A more detailed calculation yields

\bea
\eta_0(\gamma)&=&2-2\gamma, \nonumber \\
\eta_1(\gamma)&=&2-{d\over2}+2\gamma(1+{3\over M})+\gamma^2{24\over d}
(1+{13\over M}+{12\over M^2})+{\cal O}(\gamma^3), \nonumber \\
\eta_2(\gamma)&=&2-{2d\over3}-
8\gamma{3M+8\over M(M+1)}+{\cal O}(\gamma^2), \nonumber \\
\eta_3(\gamma)&=&2-{3d\over4}+6\gamma{3M^2+69M+196\over M(M+1)(M+2)}+
{\cal  O}(\gamma^2), \nonumber \\
\eta_4(\gamma)&=&2-{4d\over5}-48\gamma{15M^2+215M+636\over
M(M+1)(M+2)(M+3)}+{\cal O}(\gamma^2), \nonumber \\
\eta_5(\gamma)&=&2-{5d\over6}+80\gamma{5M^3+360M^2+4225M+12846\over
M(M+1)(M+2)(M+3)(M+4)}+{\cal O}(\gamma^2), \nonumber \\
\eta_6(\gamma)&=&2-{6d\over7}-960\gamma{35M^3+1365M^2+14308M+44364\over
M(M+1)(M+2)(M+3)(M+4)(M+5)}+{\cal O}(\gamma^2). \nonumber 
\eea

Note that the expression for $\eta_0(\gamma)$ is, in fact, exact as
presented, i.e. it does not have $\gamma^{2}$ corrections. 
One recognizes it as the line of the TFP solutions. 
Moreover, note that the order $\gamma$ contribution to
$\eta_n(\gamma)$ vanishes in the limit $M \rightarrow\infty$ (and thus
for $N\rightarrow\infty$) for $n\ge2$. This is, of course, consistent
with what we expect, namely, $\eta_n$ constant for $N=\infty$ and
$n\ge2$. We would like to stress that the expansions for $\eta_{n}(\gamma)$ 
obtained here are in a full agreement with general formula (\ref{eta-sol}) 
expressing the universality of the pattern of the curves for various $d$. 

To obtain the curves $\eta_n(\gamma)$ for arbitrary values of $\gamma$ a
numerical approach was taken. We calculated them for a wide range of 
values of $N$ (up to 1000). As an illustration let us consider the 
case $d=3$ in more detail. 
For even $n$, the curves lie in the region
delineated by the lines $\gamma=0$, $\eta=-1$ $(\gamma>0)$, and
$\eta= 2 -2\gamma$ (this is the curve $\eta_0(\gamma)$). For odd
$n$ the delineating lines are $\gamma=0$, $\eta=-1$ $(\gamma<0)$. 
Higher values of $n$ correspond to lower-lying curves.  
For $1 < N < +\infty$ the pattern of the curves is very similar to 
the the $N=1$ case, plotted in Fig.~\ref{fig:d32a}. 

\begin{figure}[ht]
\center{
\psfig{file=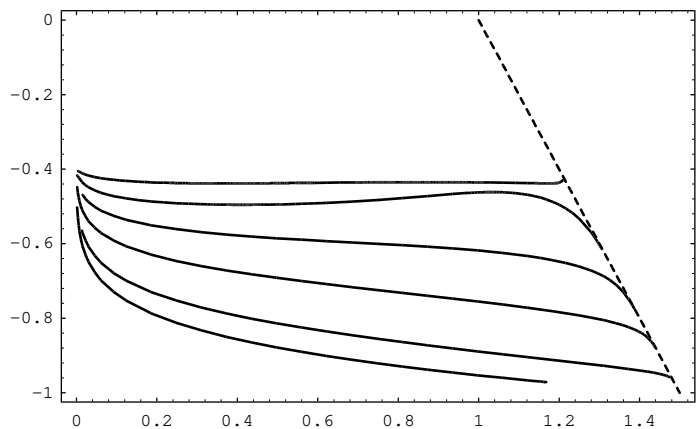,width=0.6\hsize}}
\caption{Curves in the $(\gamma,\eta)$-plane representing regular solutions
for the case $d=3$, $n=4$, with $N=1$, 2, 5, 10, 20, 50. Curves lying higher correspond to larger values of $N$. The dashed line 
is $\eta = 2 - 2\gamma$.}
\label{fig:n4}
\end{figure}

We found a one-to-one 
correspondence between the curves for various $N$ and fixed $n$. 
In Fig.\ \ref{fig:n4} we have plotted the case $n=4$, where the
$\eta_0(\gamma)$ line is also included.
We see that for increasing values of $N$ the curves lie higher.
In the $N \rightarrow \infty$ limit for $0<\gamma<s_n/2$ they approach  
the (horizontal) line $\eta=\eta_n$ and for $s_n/2 < \gamma < 3/2$ 
the line $\eta_0(\gamma)$. 
Similar curves but with lower precision were found for odd $n \geq 3$. 
\vspace{1cm}

Within the accuracy of the computation it can be concluded that as 
$N$ grows the functions 
$\eta_{n}(\gamma)$, existing for $-\infty < \gamma <  0$, 
transform to $\eta_{n}=$const, existing only for $\gamma_{*n} < \gamma < 0$. 
The latter is the limit we expect from our studies of the case $N=\infty$ 
in Sect. 3.    

\section{Discussion and conclusions}

\label{sect:concl}

We studied families of regular FP solutions of the Polchinski ERG 
equation for the $O(N)$-model in the LPA. The families are labeled by 
the integer $n \geq 1$ and are represented by curves $\eta_{n}(\gamma)$ 
in the $(\gamma,\eta)$-plane. We proved that for given $N$ the pattern of 
the curves is universal for all $d$ and described by Eq.\ (\ref{eta-sol}). 

There is a one-to-one correspondence between the curves for different $N$.  
In particular, for any $N$ 

\[ 
\eta_{n}(0) = 2-\frac{nd}{n+1},
\]
see Eq.(\ref{eta-0}), and $\eta_{n} \rightarrow 2-d$ as 
$\gamma \rightarrow -\infty$ for $n$ odd. Further properties 
of the curves were discussed in detail in Sect.~\ref{sect:Nlim}. 

As $N$ increases the $\eta_{1}(\gamma)$ curve transforms into 
the corresponding curve for $N=\infty$. The rest of the curves transform into 
the straight lines given by Eq.\ (\ref{eta-n}). 

For the $N=\infty$ case the solutions were studied analytically. 
We have shown that most of them have a 
singularity at some finite point (see Eq.\ (\ref{u1-sing})) and 
are not acceptable from the physical point of view. The condition of 
regularity imposes a relation between the parameters $\eta$ and $\gamma$.
We analyzed this condition of regularity explicitly and 
demonstrated how various classes of FP solutions appear.  

The analysis of the $N=\infty$ case, carried out in the article, gives an explicit analytical illustration of the condition which selects regular 
solutions. This enables to get an insight into the nature 
of FP solutions of the ERG equations in the LPA. Usually these 
equations are stiff, and gaining a better understanding of the 
behavior of its solutions is quite valuable. 

We would like to mention that the 
pattern of the curves resembles the pattern of eigenvalues ($\eta_{n}$) 
of an eigenvalue problem with a parameter ($\gamma$). As the parameter 
varies the levels change and various phenomena, like level crossing, 
may occur. 

The results can be extended to higher order approximations of the 
Polchinski ERG equation. In particular, 
as it was mentioned in the Introduction, the regular solutions studied here 
can be used as initial input in the iteration procedure used for solving 
the next-to-leading order approximation.   
Methods of the analysis and the obtained results can also be useful in 
studies of ERG equations of other types in scalar models, as well as in 
other classes of theories, in particular in fermionic theories. 

\section*{Acknowledgments}

We would like to thank T.R. Morris for useful discussions and valuable 
comments. 
We acknowledge financial support from the Portuguese Funda\c {c}\~ao
para a Ci\^encia e a Tecnologia (F.C.T.) under grants 
CERN/P/FIS/40108/2000, PRAXIS/2/2.1/FIS/\-286/94, POCTI/FIS/32694/2000 and 
fellowships PRAXIS/BPD/14137/97, SFRH\-/BPD/7182/2001.

\end{document}